\documentclass[a4paper, 12pt]{article}
\usepackage{amsmath}
\usepackage{graphicx}
\usepackage{hyperref}
\usepackage{geometry}
\usepackage{setspace}
\usepackage{subfigure}
\usepackage{xcolor}
\usepackage[table]{xcolor} 
\usepackage{booktabs} 
\usepackage{multirow} 
\usepackage{tabularx}
\usepackage{enumitem}
\usepackage{pgfplots}
\usepackage{float}
\usepackage{amssymb}
\usepackage{fontawesome5}
\usepackage{makecell}
\usepackage{authblk}
\usepackage{booktabs}
\usepackage{natbib}
	\setcitestyle{authoryear,round,aysep={,}}

\usepackage{subcaption}
\usepackage{tikz}
\usetikzlibrary{matrix}

\title{Empirical behavioural heterogeneity shapes the dynamics of an agent-based land use model}

\author[1]{Ronja Hotz}
\author[1]{Thomas Schmitt}
\author[1]{Calum Brown}
\author[1]{Yongchao Zeng}
\author[1,2,3]{Mark Rounsevell}

\affil[1]{Institute of Meteorology and Climate Research, Atmospheric Environmental Research (IMK-IFU), Karlsruhe Institute of Technology, 82467 Garmisch-Partenkirchen, Germany}
\affil[2]{Institute of Geography and Geo-Ecology, Karlsruhe Institute of Technology, 76131 Karlsruhe, Germany}
\affil[3]{School of Geosciences, University of Edinburgh, Drummond Street, Edinburgh EH8 9XP, UK}

\date{}

\begin{document}

\maketitle
\begin{abstract}
Land use models often represent decision-makers as homogeneous and rational, overlooking socio-psychological diversity and potentially generating rapid, coordinated land use responses that contrast with observed land use patterns. Here, we integrate an empirical typology of European forestry practitioners into an agent-based land use model by translating survey-based behavioural profiles into cognitive parameters governing endogenous decision-making processes. We distinguish five practitioner types and parametrise heterogeneous agent populations according to the empirically observed distribution of these types. Using a stylised model landscape, we compare these heterogeneous populations against a homogeneous rational-choice baseline and single-type populations. Compared with the homogeneous rational-choice baseline, heterogeneous populations dampen synchronised responses to changing ecosystem service demand, producing more gradual land use dynamics and a higher prevalence of medium intensity management, that better reflect empirical land use patterns. Our experiments reveal that similar land use patterns can emerge through different behavioural mechanisms, while the behaviour of individual decision-making types depends strongly on the population context in which they are embedded. Together, these two findings show that individual behavioural responses and land use outcomes co-evolve rather than decision-making types mapping onto fixed management practices. Explicitly representing socio-psychological diversity can therefore improve the realism and policy-relevance of agent-based land-use models by capturing feedbacks between cognition, social interactions, and emergent land use dynamics.
Empirical data may make this representation possible, but models can also use this capability to explore behavioural heterogeneity as a key source of uncertainty when data are absent.
\end{abstract}

\noindent\textbf{Keywords:}
Agent-based model, Land use, Decision-making, Behavioural heterogeneity, Typology

\section{Introduction}

Land managers’ decisions shape land use patterns and subsequently the provision of ecosystem services (ES), including timber, food and carbon sequestration, while also affecting biodiversity conservation \citep{hasan_impact_2020}. At the same time, land use contributes to greenhouse gas emissions, disrupts nutrient cycles, and can erode landscape resilience \citep{rockstrom_safe_2009}, making its careful regulation essential for maintaining Earth system stability \citep{foley_global_2005, ipbes_summary_2018}. Understanding how land use decisions are made is therefore central in the management of sustainable land use transitions.

Empirical research shows that land managers differ substantially in their values, beliefs, objectives, and susceptibility to social influence, and that these differences significantly shape land use decisions \citep{swart_meta-analyses_2023, dessart_behavioural_2019, primmer_professional_2010}. Yet these socio-psychological characteristics are rarely considered in policy instruments \citep{tiebel_small-scale_2021, dessart_behavioural_2019} and land managers are still often represented as homogeneous rational actors in large-scale land use and global assessment models \citep{brown_behavioral_2017, holman_improving_2018,fisher-vanden_evolution_2020,rubiano_rivadeneira_justice_2022}. Agent-based land use models (ABLUMs) provide a promising alternative, as they allow for the representation of heterogeneous actors and their interactions with one another and with their environment \citep{murray-rust_combining_2014}. However, even in ABLUMs, decision-making processes are often simplified and underlying socio-psychological mechanisms remain underrepresented \citep{groeneveld_theoretical_2017, huber_representation_2018}.

This is especially true for the representation of forestry practitioners' decision-making \citep{ekstrom_modelling_2024}. Forest management decisions are strongly behavioural, being shaped by personal values, traditions, and social norms in addition to economic incentives \citep{tiebel_conservation_2022, westin_forest_2023, sotirov_forest_2019, primmer_professional_2010}. At the same time, forests play a central role in meeting climate mitigation, timber production, and biodiversity targets \citep{brockerhoff_forest_2017}, making a realistic representation of forestry decision-making especially important.

To structure heterogeneous decision-making, a growing body of literature has used social surveys and empirical methods to investigate similarities and differences among land managers, grouping them into distinct behavioural typologies. A global synthesis identified recurring land use decision-maker types based on objectives, attitudes, and abilities \citep{malek_local_2019}. In Europe, diverse farmer and forest owner typologies have been derived using combinations of socio-demographic, structural, and socio-psychological variables \citep{bartkowski_typologies_2022, ficko_european_2019}. While these typologies provide valuable insights into behavioural diversity, they are often context-specific and heterogeneous in their methodological approaches, limiting their direct applicability to large-scale agent-based modelling.

To introduce structured heterogeneity into large-scale ABLUMs, a sufficient degree of abstraction is required. \citet{arneth_global_2014} proposed the concept of Agent Functional Types (AFTs) as a theoretically informed typological approach to enable scaling up agent-based models. AFTs combine roles (e.g.\ forester, farmer) with behavioural characteristics (e.g.\ risk aversion, imitation, conservatism) and are implemented as generic functional classes. Following this concept, \citet{blanco_characterising_2015} developed a generic typology of forest owners based on a meta-analysis across the developed world, which was subsequently applied in a national-scale ABLUM to define forest owner AFTs according to their objectives and associated management practices \citep{blanco_importance_2017}. 

However, in current ABLUMs that use land manager typologies, differences between types are primarily expressed through observable management behaviours, while underlying socio-psychological mechanisms guiding decision-making are not explicitly modelled. Either they remain restricted to local scales, particularly in forestry contexts \citep{ekstrom_modelling_2024}, or represent heterogeneity mainly through variations in management styles and objectives without endogenously simulating socio-psychological processes such as the formation of attitudes and social norms \citep{huber_representation_2018}. As a result, these approaches assume a direct correspondence between agents’ objectives and their management behaviours. This makes it difficult to represent intention–behaviour gaps that have been identified as significant in land use decision-making \citep{ficko_european_2019, swart_meta-analyses_2023}.

To address this gap, we build on a generic behavioural extension for ABLUMs \citep{hotz_modelling_2026} that explicitly models socio-psychological drivers of land management intensity decisions by incorporating attitudes, social norms, and behavioural inertia into a cognitive decision layer. We integrate this extension into the large-scale ABLUM framework CRAFTY (Competition for Resources between Agent Functional TYpes) \citep{murray-rust_combining_2014} and parameterise it using an empirical typology of European forestry practitioners derived from a large-scale social survey \citep{diana_feliciano_decision_2025}. 

In contrast to many existing European forest owner typologies, which primarily cluster owners based on structural characteristics or management objectives \citep{ficko_european_2019}, this typology is explicitly grounded in an explanatory behavioural theory, the Theory of Reasoned Action (TRA) \citep{fishbein_predicting_2010}. The TRA captures belief-based components of decision-making that form the foundation of the Theory of Planned Behaviour (TPB) \citep{ajzen_theory_1991}, on which the behavioural model extension is conceptually based. This strong theoretical alignment, together with the cross-country scale of the social survey spanning thirteen European countries, makes the typology particularly well suited for parameterising a cognitive decision layer in agent-based land use modelling.

Our approach explicitly distinguishes between (i) land management behaviours represented as agent functional types (e.g.\ high intensity, medium intensity, and conservation management) and (ii) decision-making types characterised by different underlying beliefs and socio-psychological dispositions. This separation allows us to represent situations in which intended objectives and realised management practices do not align.

We apply the model in a stylised landscape to isolate the effects of behavioural diversity. Agents are parameterised according to empirically observed regional distributions of forestry practitioner types derived from the European survey. By comparing simulations reflecting different European regional compositions, we examine how real-world differences in decision-making structures translate into distinct emergent land use patterns and ecosystem service outcomes. Specifically, we address the following research questions:

\begin{enumerate}
    \item How do empirically informed regional behavioural compositions shape aggregate land use dynamics and ecosystem service provision in comparison to rational choice assumptions?
    
    \item How do individual behavioural types within regional compositions contribute to emergent aggregate outcomes?
    
    \item How does the behavioural composition of the population influence the behaviour of individual behavioural types?
\end{enumerate}

\section{Methods}
\subsection{Overview}
We combine a generic behavioural model extension for land managers’ decision-making \citep{hotz_modelling_2026} with the ABLUM framework CRAFTY \citep{murray-rust_combining_2014}. In the integrated model, land managers decide on switching between different land management intensities based on
their environmental attitudes, descriptive social norms, behavioural inertia and demand-driven competitiveness.

We use a typology for European forestry practitioners\citep{diana_feliciano_decision_2025} to parameterise the cognitive decision layer, distinguishing five types with empirically grounded socio-psychological profiles. We conduct simulations in a stylised landscape and assign agents behavioural parametrisation according to observed compositions of practitioner types in different European regions. Emergent land use outcomes are compared across region-specific behavioural configurations, homogeneous rational-agent benchmarks, and single-type landscapes under contrasting ecosystem service demand scenarios.

The following subsections describe (i) the behavioural model extension and its integration into CRAFTY, (ii) the translation of the typology into model parameters, and (iii) the experimental design and evaluation procedures.

\subsection{Behavioural Model}

We build on the generic behavioural extension introduced in \citep{hotz_modelling_2026}. 
The model integrates socio-psychological drivers of land management intensity 
decisions into agent-based land use models, guided by the TPB \citep{ajzen_theory_1991} as an overarching framework.

The behavioural layer determines whether a land manager switches between land 
management intensities based on three components:

\begin{enumerate}
    \item \textbf{Environmental attitude ($A_{\alpha}$)} \\
    Represents pro-environmental ($A_{\alpha}>0$) versus productivist ($A_{\alpha}<0$) orientations and biases 
    decisions toward extensification or intensification.

    \item \textbf{Descriptive social norm ($S^{\alpha}_{AB}$)} \\
    Captures peer influence through empirical expectations within a social 
    network, based on spatial proximity and optionally additional long distance ties. Agents 
    respond to the observed prevalence of alternative management intensities 
    relative to individual critical mass thresholds. The weight of social norms compared to environmental attitudes is given by $w_{\alpha}$.

    \item \textbf{Behavioural inertia ($\lambda_{\alpha}$)} \\
    Represents limited perceived behavioural control and increases the resistance to change with the 
    magnitude of the intended intensity change, thereby stabilising existing 
    practices.
\end{enumerate}

These components are aggregated into a composite behavioural influence score, which is transformed via a logistic function into a dynamic giving-in threshold. This threshold represents a barrier that must be overcome for a new land management practice to be adopted and serves as the link to the host land use model. Its maximum value is given by $L_{\alpha}$, defining the importance given to socio-psychological factors compared to competitiveness. In the case of CRAFTY, a transition occurs when the utility advantage of a competing management intensity exceeds this threshold. While CRAFTY applies this threshold mechanism to all land use transitions, the behavioural model is limited to transitions between land management intensities and does not represent transitions between land management categories such as forestry and agriculture.

The behavioural extension is generic and modular, operating independently of the specific land allocation mechanics of the host model; see \citep{hotz_modelling_2026} for the full formalisation and equations.

\subsection{CRAFTY Modelling Framework}

The behavioural extension is integrated into \textsc{CRAFTY}, an 
ABLUM framework designed to simulate spatial 
land use dynamics under competing ecosystem service demands 
\citep{murray-rust_combining_2014,brown_societal_2019,brown_agentbased_2022}.

In CRAFTY, the landscape is represented as a spatial grid of cells. Each cell is managed by an agent belonging to an AFT. AFTs define production functions and sensitivities to local biophysical and socio-economic capitals. In our application, they represent land management practices, which are distinct from the behavioural typology used for the parametrisation of the decision-making. Agents compete for land based on their ability to meet societal ES demand.
    Land use change occurs when a competitor's utility exceeds that of 
    the incumbent by more than a giving-in threshold.

With the behavioural extension, this threshold is endogenously calculated. Behavioural parameters 
are stored as cell-level properties, enabling spatially heterogeneous decision-making.

\subsection{Behavioural Typology and Parameterisation}

To ground the behavioural parametrisation of our model in empirical data, we draw on a typology of forestry practitioners \citep{diana_feliciano_decision_2025}. This typology was developed from a large-scale social survey conducted in thirteen European countries. It categorises forestry practitioners into distinct groups based on their underlying behavioural, normative, and control beliefs following the TRA and distinguishes Environmentally Conscious Passives, Environmental Implementers, Traditionalists, Maximisers, and Social Satisfiers. We use this typology to parameterise agents' decision-making processes by assigning distinct behavioural model parameter combinations (Table~\ref{tab:typology}). In addition, survey-based regional distributions of these practitioner types \citep{diana_feliciano_decision_2025} are used to initialise heterogeneous agent populations in the simulation experiments, enabling us to incorporate empirically grounded behavioural diversity into the model and to examine how differences in regional decision-making structures shape land use outcomes. 

The empirical typology does not provide direct numerical estimates of the cognitive
parameters used in the model. We therefore parameterised the behavioural types through
a structured interpretative translation of the ordinal empirical indicators and qualitative
type descriptions. The empirical indicators describe the relative direction and strength of
different behavioural characteristics, ranging from strongly negative to strongly positive
(Table~\ref{tab:typology_matrix}). They were therefore treated as qualitative evidence about relative differences
between types, rather than as statistically estimated effect sizes. To avoid implying
unsupported numerical precision, parameter values were assigned using a limited set of
discrete levels representing ordinal differences between behavioural types. The translation
focused on those empirical indicators that correspond conceptually to mechanisms
represented in the behavioural decision layer. Table~\ref{tab:indicator_parameter_mapping}
summarises which empirical information was used for each model parameter.
Further details about the parametrisation of the typology are provided in Appendix \ref{appendix_C}.

\begin{table}[H]
\centering
\begingroup
\footnotesize
\renewcommand{\arraystretch}{1.08}
\setlength{\tabcolsep}{4pt}
\rowcolors{2}{gray!12}{white}

\begin{tabularx}{\textwidth}{
    >{\raggedright\arraybackslash}X
    c c c c c c
}
\toprule
\textbf{Behavioural Type} & \textbf{$L_{\alpha}$} & \textbf{$w_{\alpha}$} & \textbf{$A_{\alpha}$} & \textbf{$\lambda_{\alpha}$} & \textbf{North} & \textbf{Southwest} \\
\midrule

\faBed\ \faLeaf\ \textbf{Environmentally Conscious Passives}\par
\emph{Low interest in personal utility; limited reliance on social norms or market mechanisms; strong environmental values.}
& 1 & 0.25 & 0.5 & 0.25 & 17.20\% & 2.82\% \\

\faTools\ \faLeaf\ \textbf{Environmental Implementers}\par
\emph{Strong environmental values; not influenced by social norms; low importance of income, but value market mechanisms.}
& 0.75 & 0 & 0.75 & 0 & 13.14\% & 24.65\% \\

\faHome\ \textbf{Traditionalists}\par
\emph{Value environmental objectives and social norms (e.g. local traditions); low importance of income.}
& 0.8 & 0.5 & 0.5 & 0.25 & 16.01\% & 4.93\% \\

\faDollarSign\ \textbf{Maximisers}\par
\emph{Low environmental concern; strong emphasis on income and market mechanisms.}
& 0.25 & 0.1 & -1 & 0 & 20.91\% & 16.90\% \\

\faUsers\ \textbf{Social Satisfiers}\par
\emph{Strongly influenced by social norms; ambivalent towards income and environmental objectives.}
& 0.6 & 0.75 & 0 & 0 & 32.74\% & 50.70\% \\

\textbf{Baseline: Rational Choice}\par
\emph{Rational decision-making; no behavioural extension.}
& 0 & None & None & None & -- & -- \\

\bottomrule
\end{tabularx}

\caption{Behavioural typology of forestry practitioners based on \citet{diana_feliciano_decision_2025}, including model parametrisation and regional distribution of behavioural types in North and Southwest Europe. North and Southwest represent the percentage of observed survey respondents within the typology group in the two regions. Parameter ranges are defined as $L_{\alpha}\in[0,1]$, $w_{\alpha}\in[0,1]$, $A_{\alpha}\in[-1,1]$, and $\lambda_{\alpha}\in[0,1]$.}
\label{tab:typology}
\endgroup
\end{table}

\begin{table}[htbp]
\centering
\caption{Empirical basis for the behavioural parameterisation following \citet{diana_feliciano_decision_2025}.}
\label{tab:indicator_parameter_mapping}
\begin{tabular}{p{0.35\textwidth} p{0.55\textwidth}}
\toprule
\textbf{Model parameter} & \textbf{Empirical basis} \\
\midrule

Upper limit of the giving-in threshold 
(\(L_{\alpha}\)) 
& 
Other factors relative to income objectives and market mechanisms \\

\addlinespace

Weight of social norm relative to attitude 
(\(w_{\alpha}\)) 
& 
Society indicator relative to the attitude component, with the latter derived from regulating ecosystem service objectives relative to income objectives \\

\addlinespace

Environmental attitude 
(\(A_{\alpha}\)) 
& 
Regulating ecosystem service objectives relative to income objectives \\

\addlinespace

Behavioural inertia coefficient
(\(\lambda_{\alpha}\)) 
& 
Qualitative type narratives \\

\bottomrule
\end{tabular}
\end{table}

\subsection{Experimental Design}

\subsubsection{General Model Configuration}

We adopt the modelling environment introduced in \citet{hotz_modelling_2026}, embedding the behavioural decision layer within a simplified implementation of CRAFTY in NetLogo \citep{wilensky_netlogo_1999}. The model is applied in a stylised landscape consisting of a 101 × 101 grid with heterogeneous distributions of productive and natural capital (Figure \ref{fig:capitals}). This landscape is used to isolate behavioural effects. By abstracting away real-world geographic complexities, edge effects, and non-controllable environmental factors, the setup functions as a controlled experimental environment where land-use dynamics can be attributed strictly to agent behaviour rather than spatial noise. It does not correspond to the European regions used for behavioural parametrisation.  

We consider three AFTs corresponding to land-management intensities—high intensity land management, medium intensity land management, and conservation—which differ in their sensitivities to productive and natural capital through linear production functions for material and non-material ES provision (Table \ref{tab:AFTs}). They are defined independently of agents’ decision-making characteristics. At initialisation, AFTs are randomly distributed across the landscape with equal shares, while decision-making types are assigned independently according to the proportions specified in each behavioural experiment. Capitals and demand levels remain static throughout the simulation. In each simulation step, 5\% of cells are randomly selected for potential management transitions, and simulations run until a steady or quasi-steady state is reached.

\begin{table}[ht!]
\centering
\caption{Sensitivity of agent functional types to productive and natural capital.}
\label{tab:AFTs}
\begin{tabular}{lcc}
\hline
\textbf{Agent Functional Type} & \textbf{Productive Capital} & \textbf{Natural Capital} \\
\hline
High Intensity      & 1.0 & 0.0 \\
Medium Intensity   & 0.5 & 0.5 \\
Conservation       & 0.0 & 1.0 \\
\hline
\end{tabular}
\end{table}
\subsubsection{Behavioural Composition Experiments}
We conduct three sets of experiments: (i) simulations parameterised with region-specific distributions of forestry practitioner types, (ii) simulations with homogeneous populations composed of a single behavioural type, and (iii) a baseline simulation with homogeneous rational agents that respond only to marginal demand levels.

To represent regional variation, we use survey-based distributions of behavioural types from the two European regions that differ most strongly in their composition, namely Northern Europe and Southwestern Europe (Table~\ref{tab:typology}). Northern Europe comprises survey respondents from Finland, Sweden and Latvia, whereas Southwestern Europe comprises respondents from Spain and Italy. These regional compositions are implemented in separate simulation runs using the same landscape. In each run, the respective behavioural composition is assigned across the entire landscape rather than to distinct spatial subregions. The simulated landscape is not intended to reproduce the real geography of either region. The experiments should therefore be understood as controlled scenario analyses, asking how land use dynamics would unfold if agents exhibited the behavioural profiles observed among forestry practitioners in Northern or Southwestern Europe. By holding the landscape constant, this design isolates the effects of behavioural structure from confounding differences in environmental context. Agents are initialised stochastically according to the North and Southwest regional behavioural shares reported in Table~\ref{tab:typology}; realised proportions match the target distributions closely, with only minor deviations due to random assignment on a finite grid.

Beyond comparing regional mixtures, we also analyse the behaviour of each practitioner type within heterogeneous populations. That is, we examine how Environmental Implementers, Maximisers, Traditionalists, Environmentally Conscious Passives, and Social Satisfiers behave when interacting with other types in the same landscape, and compare this with their behaviour in homogeneous simulations where they interact only with agents of their own type. This makes it possible to assess both the aggregate effects of region-specific behavioural compositions and the extent to which the behaviour of each type depends on population context.

The rational-agent baseline serves as a benchmark against which we assess the effects of both behavioural decision-making and behavioural heterogeneity. The rational-agent baseline does not assume perfect foresight or global knowledge. Agents respond to marginal ecosystem service demand signals based on the competitiveness of alternative land management intensities in their own cell, but do not anticipate the simultaneous decisions of other agents elsewhere in the landscape.

\subsubsection{Ecosystem Service Demand Scenarios}

We consider two aggregated ES categories, material and non-material ES. This deliberately stylised representation is consistent with the simplified landscape and demand setup and is intended to isolate the effects of socio-psychological heterogeneity in decision-making rather than to provide a detailed representation of ecosystem service bundles. Material ES can be interpreted as provisioning services such as timber production, whereas non-material ES represent services such as recreation as well as outcomes such as biodiversity conservation.

We examine land use dynamics under three alternative ES demand scenarios, defined by different relative levels of demand for these two categories (Table~\ref{tab:es_scenarios}): high non-material demand, balanced demand, and high material demand.

\begin{table}[htbp]
\centering
\caption{Stylised ecosystem service (ES) demand scenarios.}
\label{tab:es_scenarios}
\begin{tabularx}{\linewidth}{lccX}
\toprule
Scenario Name & Material & Non-material & Illustrative interpretation \\
\midrule
Non-material focused & Low & High & Emphasis on biodiversity and recreation \\
Balanced & Medium & Medium & Balanced emphasis on timber and non-material ES \\
Material focused & High & Low & Emphasis on timber production \\
\bottomrule
\end{tabularx}
\end{table}

\section{Results}

\subsection{Effects of behavioural composition}

\begin{figure}[!h]
\centering

\subfigure[Non-material focused demand]{
    \includegraphics[width=0.47\linewidth]
    {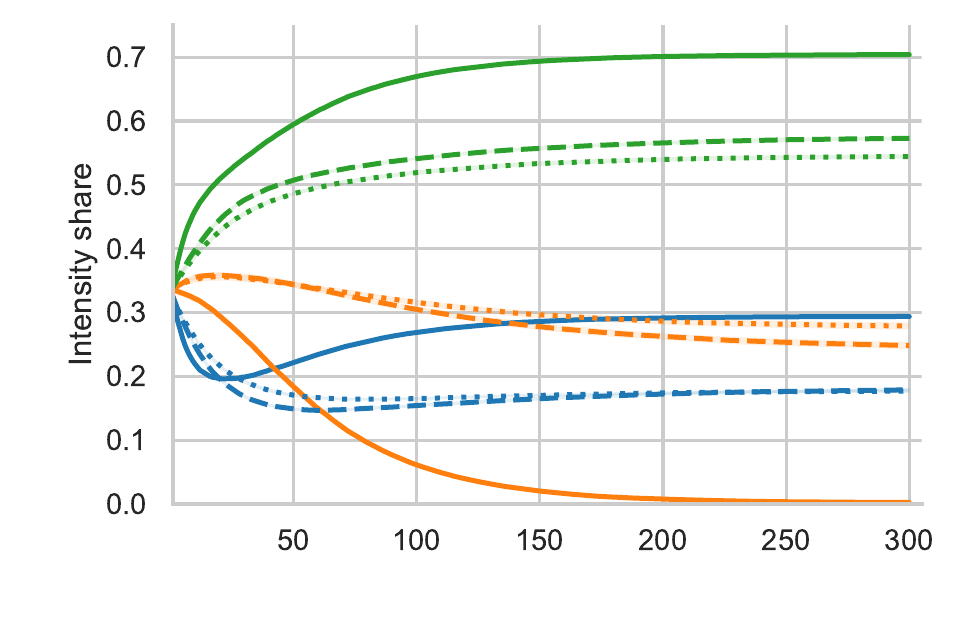}
}
\hfill
\subfigure[Balanced demand]{
    \includegraphics[width=0.47\linewidth]
    {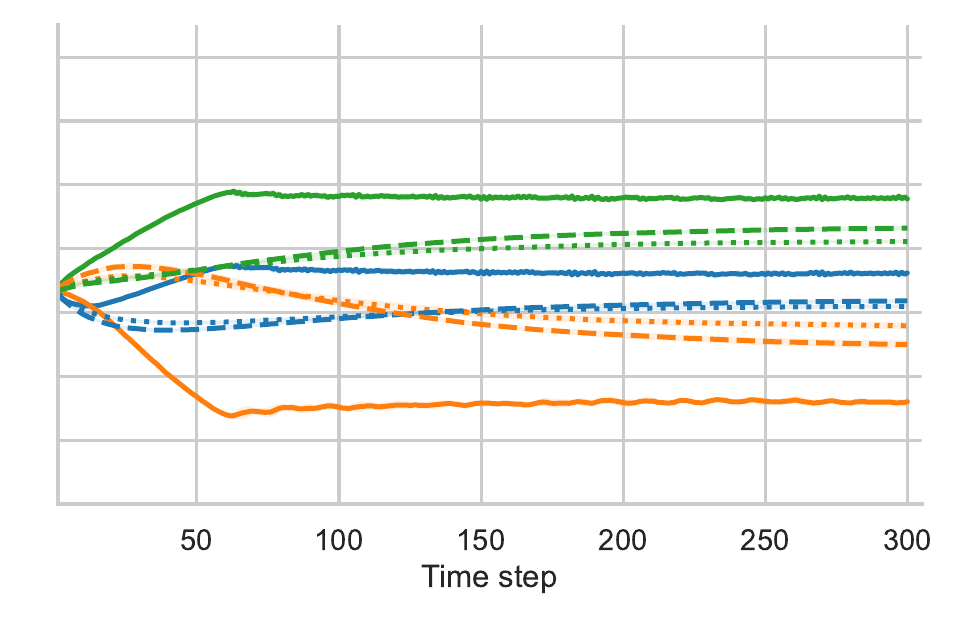}
}

\vspace{0.15cm}

\begin{minipage}{0.68\linewidth}
    \centering
    \subfigure[Material focused demand]{
        \includegraphics[width=\linewidth]
        {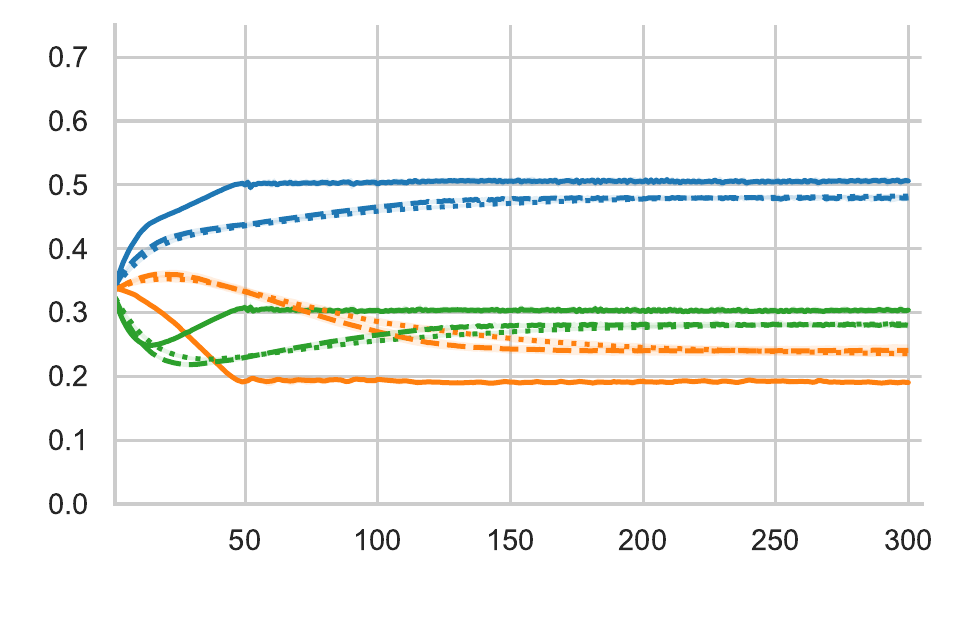}
    }
\end{minipage}
\hfill
\begin{minipage}{0.28\linewidth}
    \centering
    \vspace{0.5cm}
    \includegraphics[width=\linewidth]
    {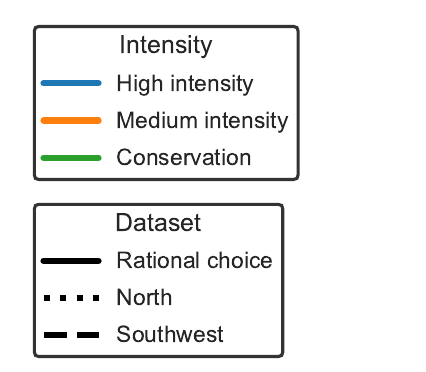}
\end{minipage}

\caption{
Aggregate land management intensity shares under different ecosystem service demand scenarios comparing North, Southwest, and rational-choice agents.
}

\label{fig:intensity_shares}

\end{figure}
Behavioural compositions representing real-world decision-making exhibit consistently higher shares of medium intensity land management than rational choice assumptions (Figure \ref{fig:intensity_shares}). Conversely, rational choice leads to a higher prevalence of extreme management strategies, i.e. high intensity and conservation-oriented land management. In addition, behavioural diversity exerts a dampening effect on system dynamics. Compared to rational choice, heterogeneous behavioural compositions lead to slower system responses, reflected in more gradual changes in land management practices after initialisation and reduced oscillatory behaviour (Figure \ref{fig:intensity_shares}). In contrast, rational choice agents respond more synchronously to marginal demand changes, producing small but persistent fluctuations in land management intensity shares and ecosystem service provision. 

\begin{figure}[!h]
\centering

\subfigure[Non-material focused demand]{
    \includegraphics[width=0.47\linewidth]
    {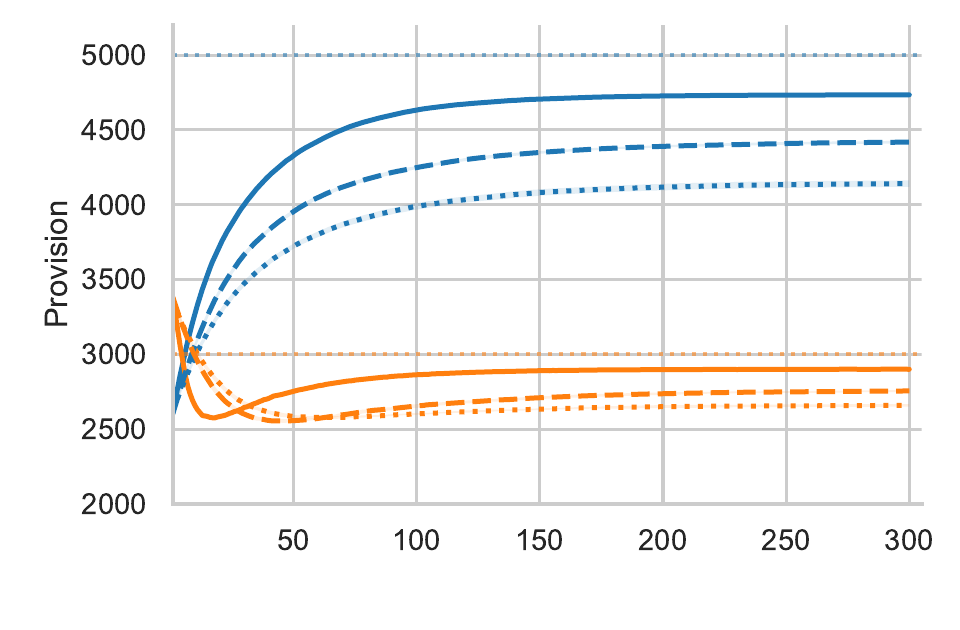}
}
\hfill
\subfigure[Balanced demand]{
    \includegraphics[width=0.47\linewidth]
    {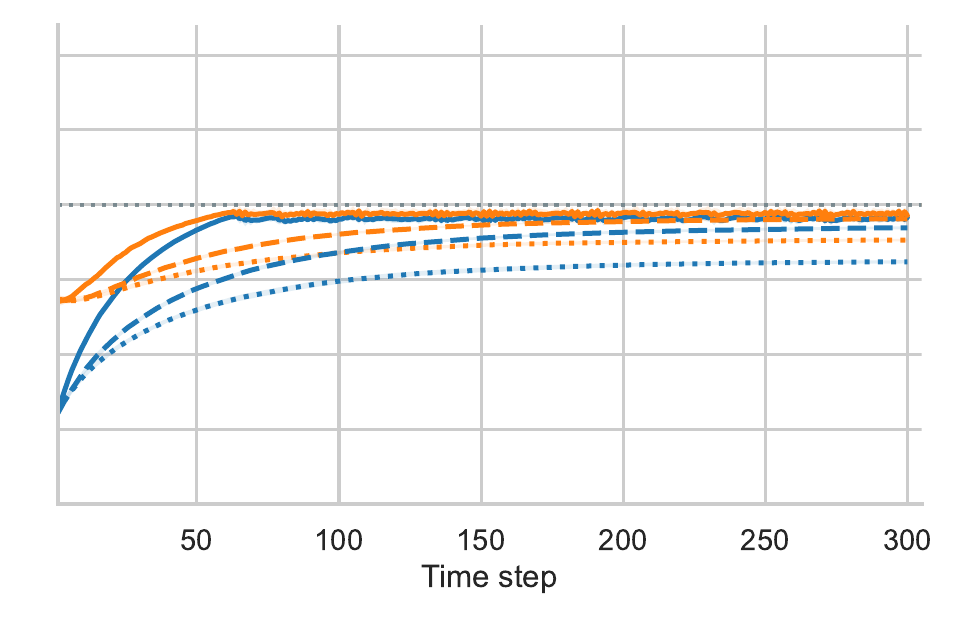}
}

\vspace{0.15cm}

\begin{minipage}{0.68\linewidth}
    \centering
    \subfigure[Material focused demand]{
        \includegraphics[width=\linewidth]
        {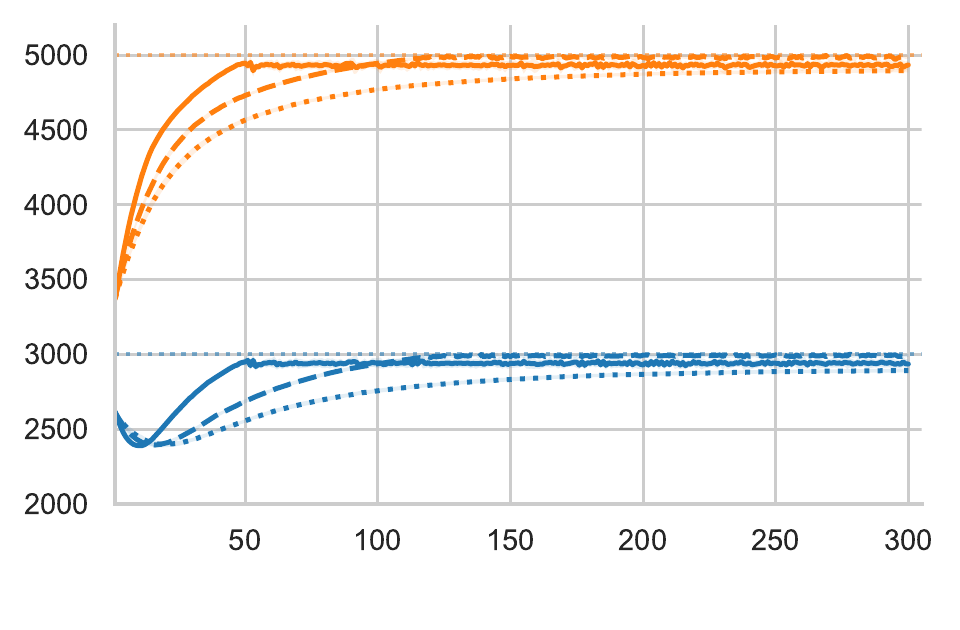}
    }
\end{minipage}
\hfill
\begin{minipage}{0.28\linewidth}
    \centering
    \vspace{0.5cm}
    \includegraphics[width=\linewidth]
    {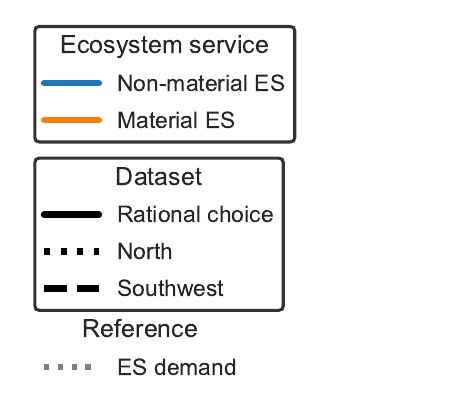}
\end{minipage}

\caption{
Aggregate ecosystem service provision under different ecosystem service demand scenarios comparing North, Southwest, and rational-choice agents. Horizontal dotted lines indicate ecosystem service demand levels.
}

\label{fig:es_supply}

\end{figure}

Differences in behavioural composition translate into distinct ecosystem service outcomes. Under the material focused demand scenario, only the Southwest behavioural composition meets demand levels (Figure \ref{fig:es_supply}). In this case, it provides the highest supply of both ecosystem services, outperforming rational choice and matching demand. Rational choice agents, in contrast, never exactly meet demand levels and instead display persistent overshooting and undersupplying dynamics. For the remaining demand scenarios, none of the behavioural compositions fully meets demand levels. In these cases, rational choice produces the highest total supply for both ecosystem services. Across all scenarios, the Southwest composition consistently provides higher levels of ecosystem services than the North (Figure \ref{fig:es_supply}).

\begin{figure}[!h]
\centering

\subfigure[Southwest]{
  \includegraphics[width=0.23\linewidth]
  {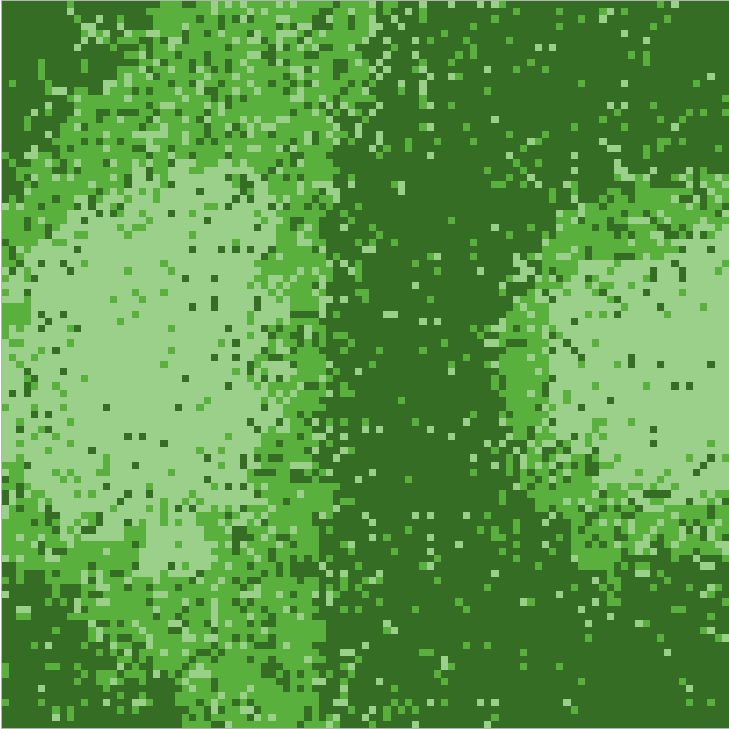}
  \label{fig:southwest}
}
\hspace{-0.22cm}
\subfigure[North]{
  \includegraphics[width=0.23\linewidth]
  {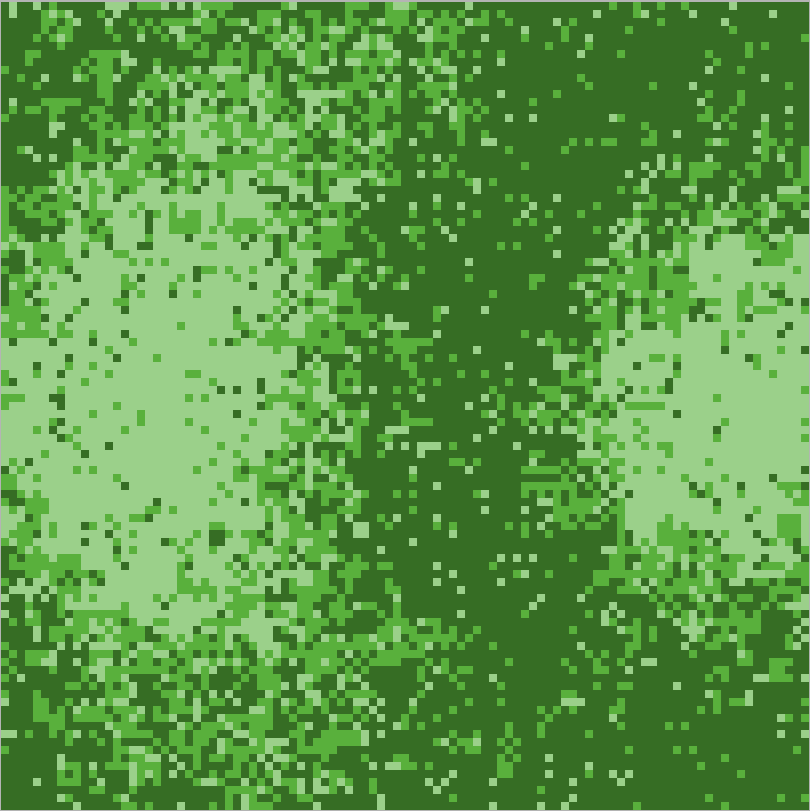}
  \label{fig:north}
}
\hspace{-0.22cm}
\subfigure[Rational Choice]{
  \includegraphics[width=0.23\linewidth]
  {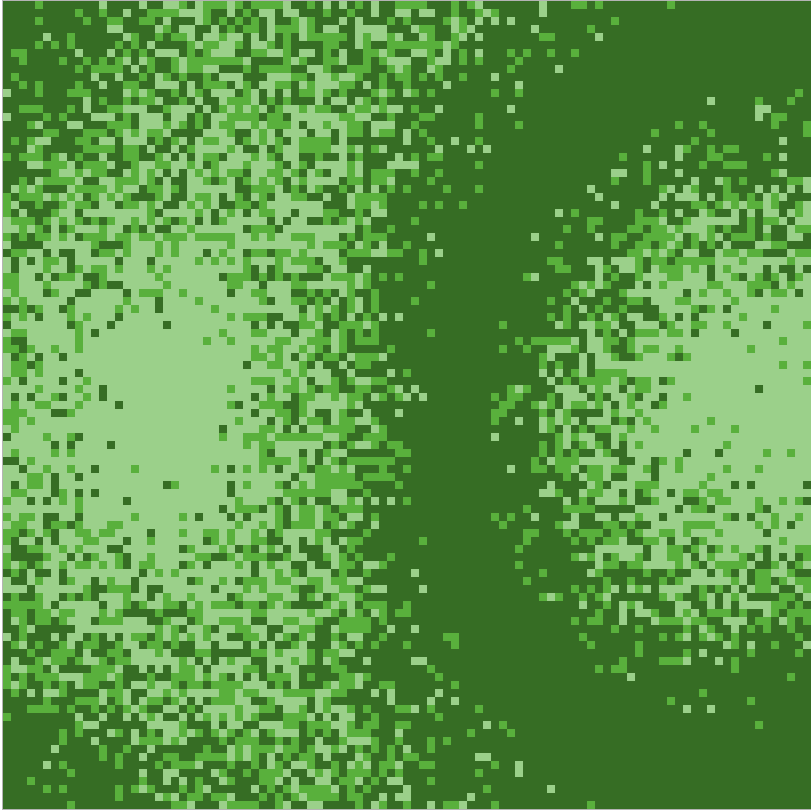}
  \label{fig:rational}
}
\hspace{-0.22cm}
\subfigure[Social Satisfiers]{
  \includegraphics[width=0.23\linewidth]
  {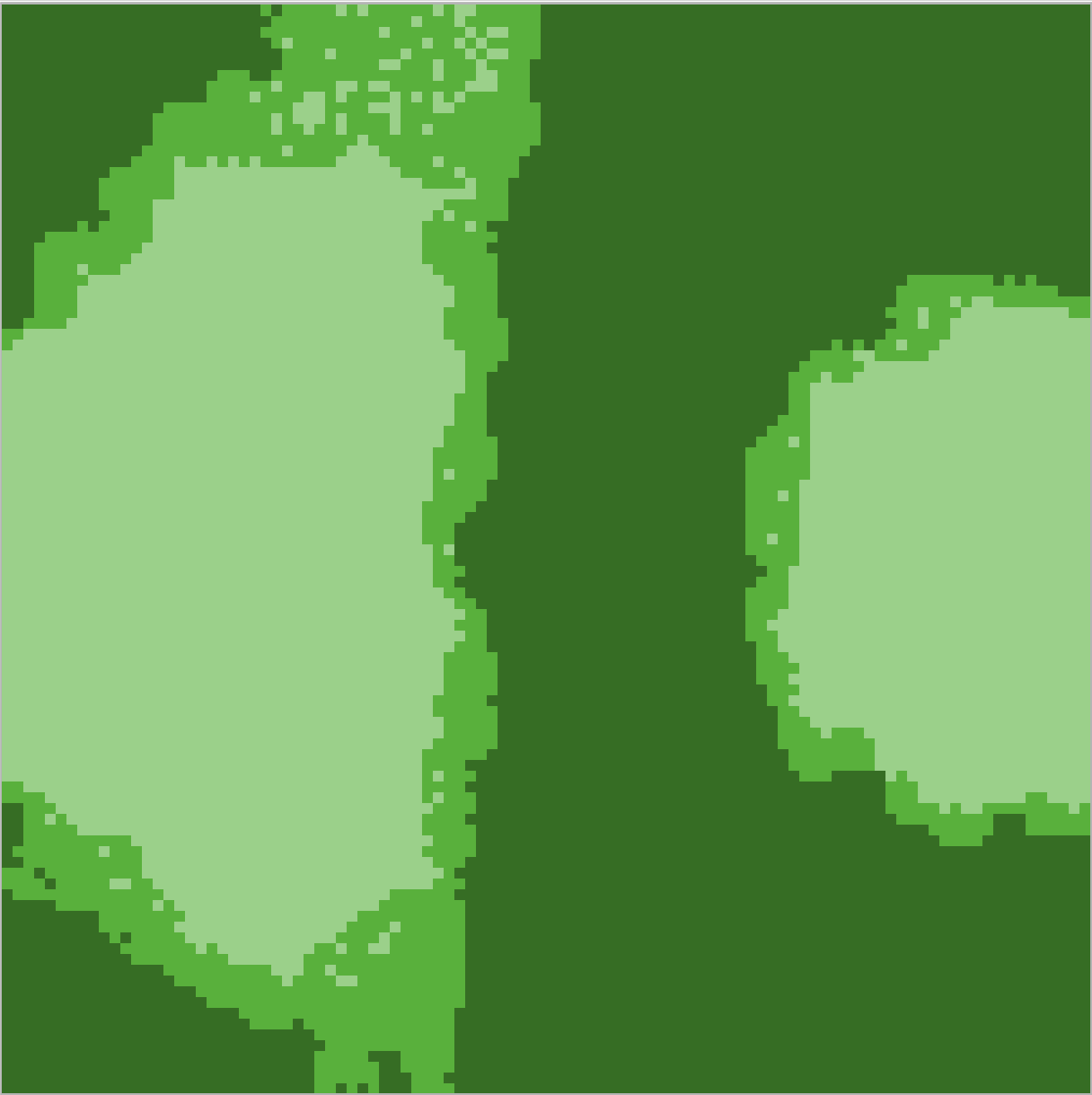}
  \label{fig:social_satisfiers}
}

\caption{
Comparison of land-use maps at the final time step for different behavioural compositions under material focused demand. Regional behavioural compositions (Southwest and North) are compared with rational-choice agents and a homogeneous population of Social Satisfiers. Colours indicate land-management intensity, with dark green representing high intensity land management, medium green medium-intensity land management, and bright green conservation.
}

\label{fig:lu_maps}

\end{figure}

Behavioural composition also affects spatial patterns of land management (Figure \ref{fig:lu_maps}). Under rational choice, areas with high productive capital exhibit more homogeneous adoption of high intensity land management compared to behavioural compositions. In contrast, the Southwest composition shows increased clustering of medium intensity land management, particularly in transitional zones. In the North composition, land management practices are more fragmented.

\subsection{Sensitivity of behavioural types to population context}

\begin{figure}[h!]
    \centering
    \includegraphics[width=1\linewidth]{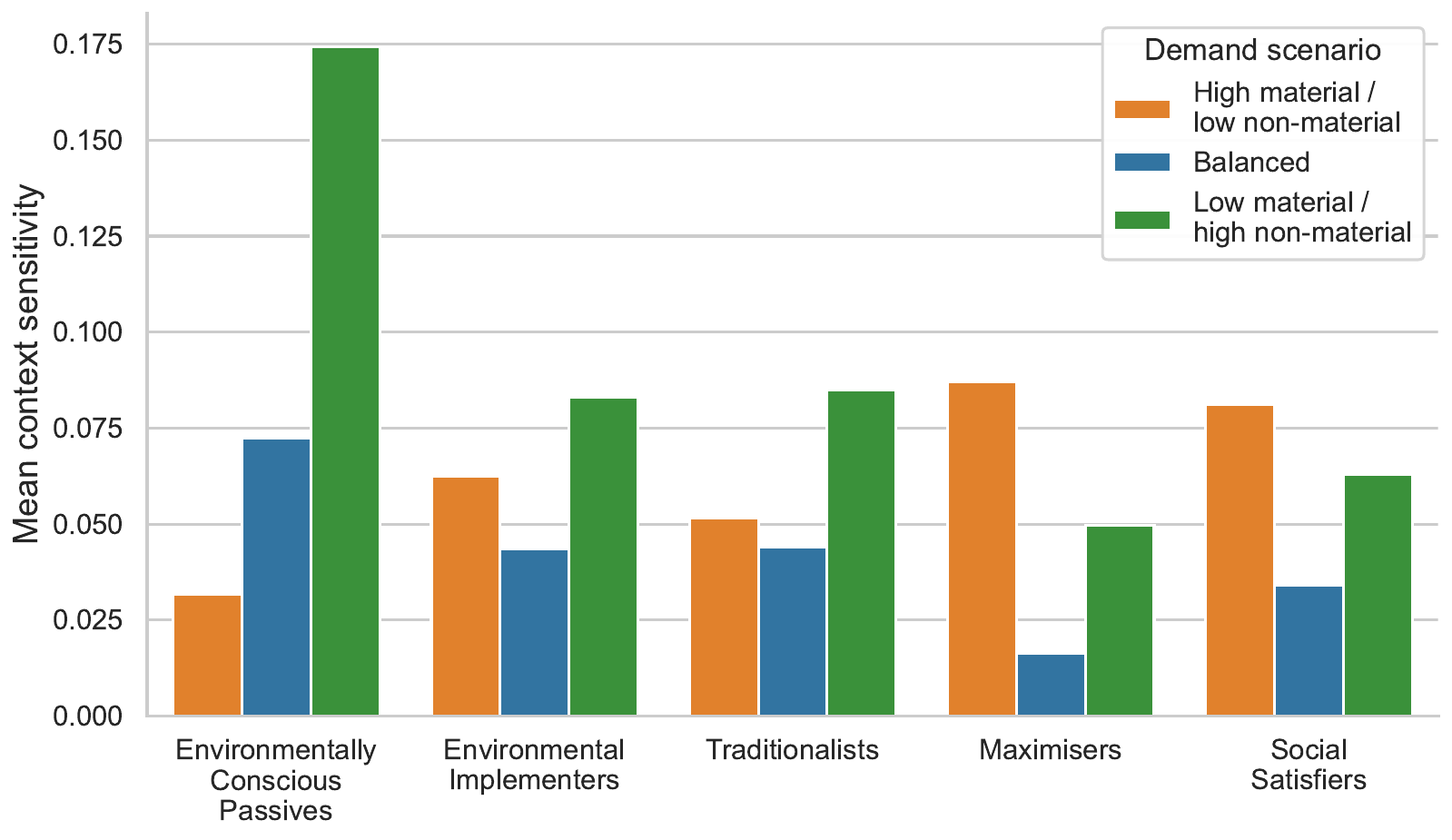}
    \caption{
Mean context sensitivity of behavioural types across ecosystem service demand scenarios. Context sensitivity is measured as the pairwise distance in land management intensity shares between behavioural compositions under identical ecosystem service demand conditions. Higher values represent stronger sensitivity to social context.
}
    \label{fig:conext_sensitivity}
\end{figure}

\begin{figure}[!ht]
\centering

\includegraphics[width=0.85\linewidth]
{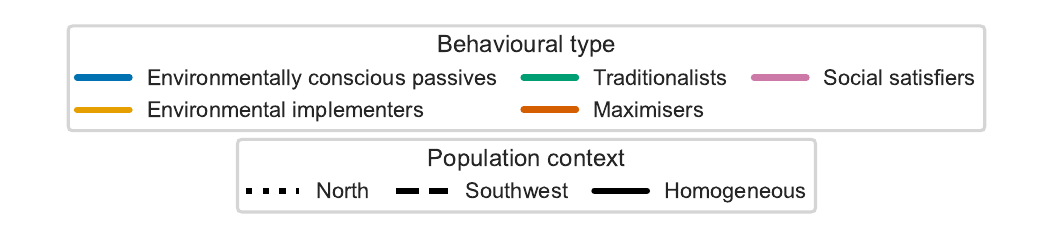}

\vspace{0.15cm}

\subfigure[Medium intensity share under material focused demand]{
    \includegraphics[width=0.48\linewidth]
    {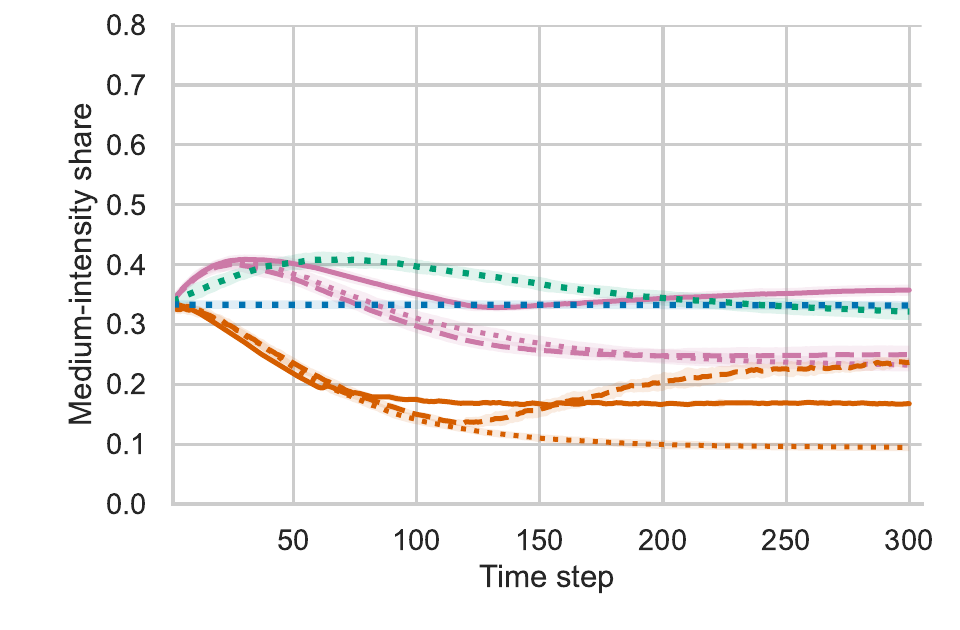}
    \label{fig:selected_type_dynamics_mif}
}
\hspace{-0.18cm}
\subfigure[Conservation share non-material focused demand]{
    \includegraphics[width=0.48\linewidth]
    {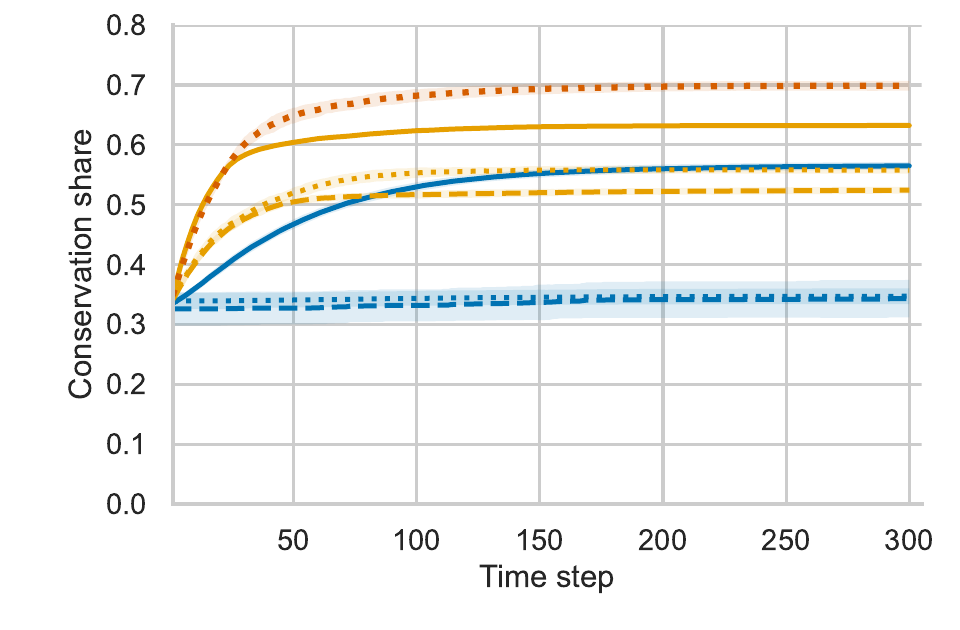}
    \label{fig:selected_type_dynamics_lif}
}

\caption{
Selected behavioural-type dynamics under contrasting ecosystem service demand scenarios. Panel (a) shows medium-intensity land management under high material/low non-material demand for social satisfiers and maximisers across all population contexts, and for traditionalists and environmentally conscious passives in the North context. Panel (b) shows conservation-oriented land management under low material/high non-material demand for environmentally conscious passives and environmental implementers across all population contexts, and for maximisers in the North context. Shaded areas indicate standard deviation across simulation runs. 
}

\label{fig:selected_type_dynamics}

\end{figure}

The behaviour of individual decision-making types depends strongly on population context and differs substantially between homogeneous settings and heterogeneous behavioural compositions. However, behavioural types vary considerably in the extent and temporal dynamics of their sensitivity to surrounding agents.
Environmentally Conscious Passives exhibit the highest overall sensitivity to population context, particularly under non-material focused demand (Figure \ref{fig:conext_sensitivity}). In heterogeneous populations, their intensity shares remain comparatively stable over time, whereas homogeneous settings produce substantially stronger behavioural dynamics (Figure \ref{fig:selected_type_dynamics_lif}).
Maximisers show the strongest context-dependent differences under material focused demand (Figure \ref{fig:conext_sensitivity}). In this scenario, their long-term adoption of medium intensity land management is substantially higher in the Southwest composition than in the North and homogeneous configurations (Figure \ref{fig:selected_type_dynamics_mif}). However, across demand scenarios, Maximisers display the lowest overall context sensitivity. Under balanced demand conditions, differences between behavioural configurations remain comparatively small (Figure \ref{fig:conext_sensitivity}).
Environmental Implementers display pronounced short-term sensitivity to population context immediately after initialisation. In homogeneous settings, they initially adopt substantially higher levels of conservation-oriented land management than in heterogeneous compositions (Figure \ref{fig:selected_type_dynamics_lif}).
In contrast, Social Satisfiers initially behave similarly across behavioural configurations, indicating comparatively low short-term context sensitivity. Over time, however, their behaviour diverges more strongly between compositions, particularly under material focused demand conditions, suggesting increasing long-term sensitivity to surrounding behavioural types (Figure \ref{fig:selected_type_dynamics_mif}).
Overall, these results indicate that behavioural types differ not only in their intrinsic decision-making characteristics, but also in the extent to which their behaviour is shaped by surrounding agents. This generates distinct context sensitivity profiles across behavioural types and demand conditions. 



\subsection{Role of behavioural types in shaping emergent patterns}

\begin{figure}[!ht]
\centering

\includegraphics[width=0.85\linewidth]
{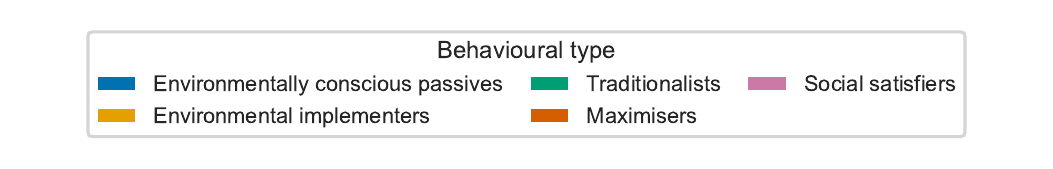}

\vspace{0.15cm}

\subfigure[North]{
    \includegraphics[width=0.48\linewidth]
    {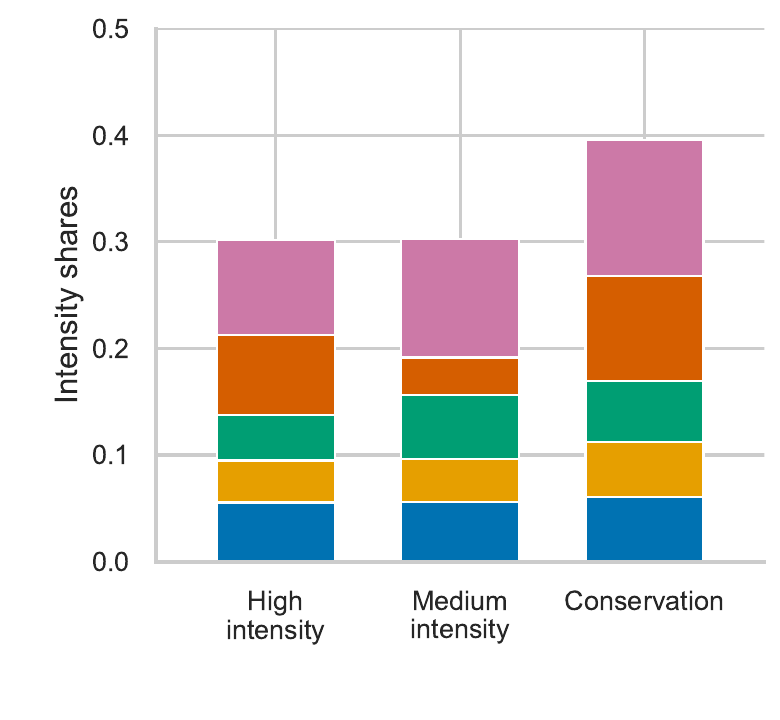}
}
\hspace{-0.25cm}
\subfigure[Southwest]{
    \includegraphics[width=0.48\linewidth]
    {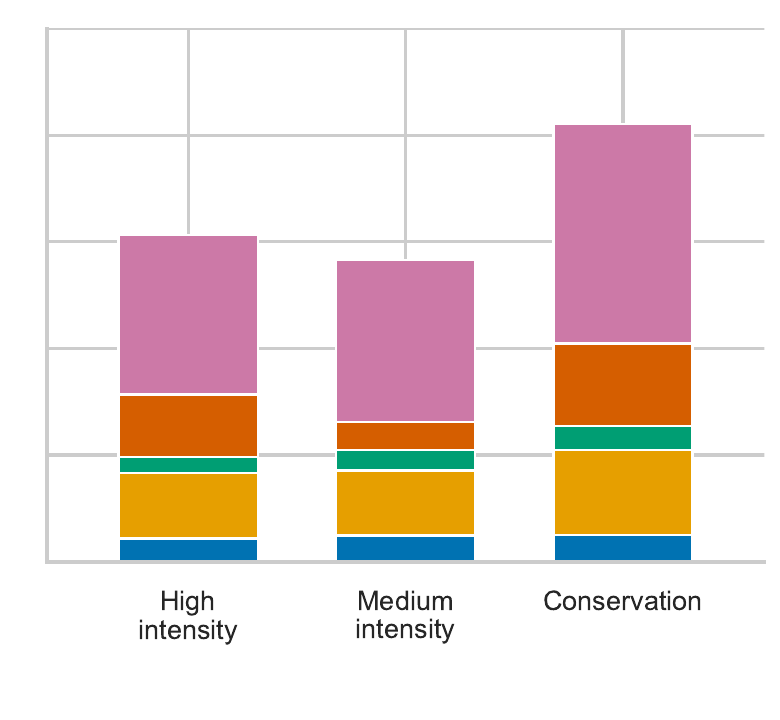}
}

\caption{
Contributions of behavioural types to aggregate land management intensity shares under balanced ecosystem service demand. Stacked bars show the contribution of each behavioural type to high-intensity, medium-intensity, and conservation-oriented land management, averaged over the final 50 time steps.
}

\label{fig:stacked_intensity_contributions}

\end{figure}

\begin{figure}[!ht]
\centering

\includegraphics[width=0.85\linewidth]
{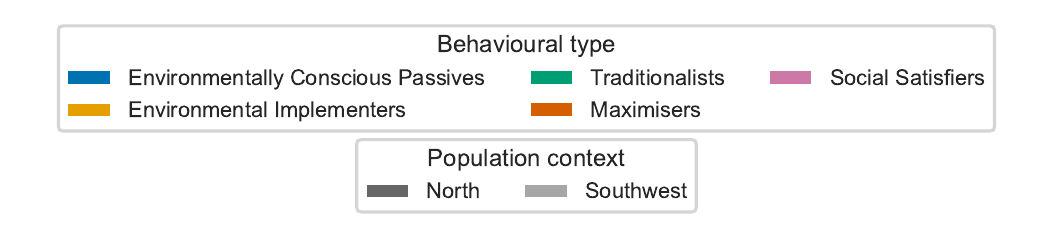}

\vspace{0.15cm}

\subfigure[Material ecosystem service provision]{
    \includegraphics[width=0.48\linewidth]
    {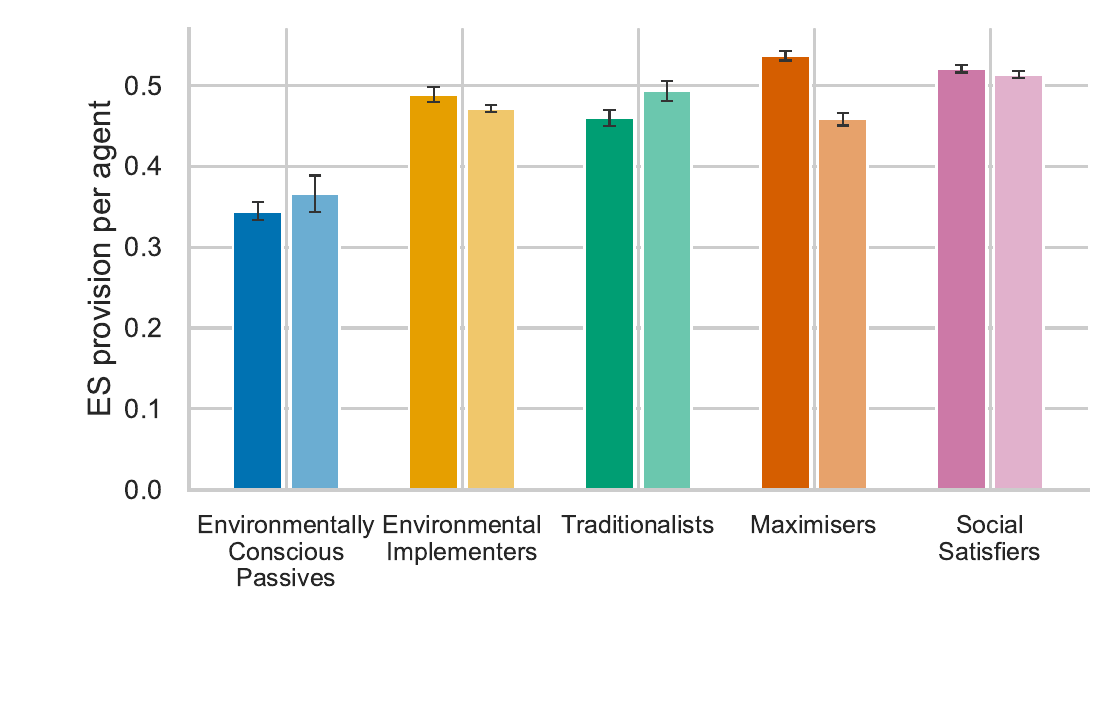}
}
\hspace{-0.25cm}
\subfigure[Non-material ecosystem service provision]{
    \includegraphics[width=0.48\linewidth]
    {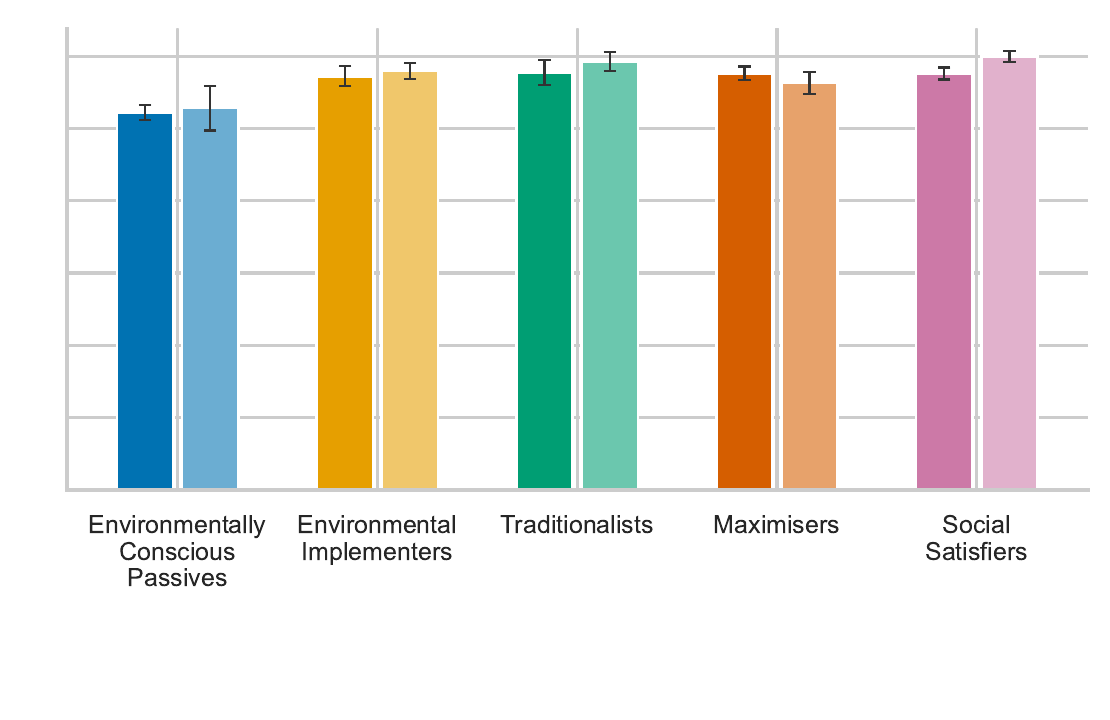}
}

\caption{
Material and non-material ecosystem service provision per agent by behavioural type under material focused demand. Bars show mean provision averaged over the final ten time steps. Error bars indicate standard deviation.
}

\label{fig:es_per_type}

\end{figure}

Emergent outcomes in the Southwest composition are strongly shaped by Social Satisfiers (Figure \ref{fig:stacked_intensity_contributions}) due to their high population share (50.7\%). Their influence on aggregate land use patterns and ecosystem service provision depends both on their abundance and on their behaviour under the respective behavioural composition and demand setting. For instance, under material focused demand, Social Satisfiers contribute strongly to the ability of the Southwest composition to meet demand, as they provide the highest levels of both material and non-material ecosystem services per agent among all behavioural types in addition to being the most abundant type (Figure \ref{fig:es_per_type}).

The comparatively high share of medium intensity land management in the Southwest composition relative to rational choice is largely driven by the strong contribution of Social Satisfiers and Environmental Implementers (Figure \ref{fig:stacked_intensity_contributions}). However, this mainly reflects their high abundance rather than consistently high medium intensity shares within each type. Social Satisfiers initially exhibit the highest medium intensity shares among all behavioural types, but these decline after approximately 40 time steps and eventually fall below those of Traditionalists and Environmentally Conscious Passives (Figure \ref{fig:selected_type_dynamics_mif}). Consequently, the Southwest composition exhibits higher medium intensity shares than the North composition during the early stages of the simulation, but lower shares in later stages. In general, Social Satisfiers show particularly rapid short-term behavioural adjustments. The high abundance of Social Satisfiers in the Southwest composition is also associated with the emergence of clusters of medium intensity land management in transitional zones (Figure \ref{fig:lu_maps}).

In contrast, medium intensity land management in the North composition emerges from more balanced contributions across behavioural types (Figure \ref{fig:stacked_intensity_contributions}). In addition to Social Satisfiers, Environmentally Conscious Passives and Traditionalists contribute substantially to aggregate medium intensity shares. These types exhibit the highest medium intensity shares within type for this regional composition (Figure \ref{fig:selected_type_dynamics_mif}) and are more abundant in the North than in the Southwest composition, supporting the persistence of medium intensity land management over time. 
The comparatively lower ecosystem service provision in the North composition is additionally associated with the higher abundance of Environmentally Conscious Passives (Figure \ref{fig:es_per_type}), whose stronger behavioural inertia reduces responsiveness to demand signals.

More generally, all behavioural types except Maximisers exhibit substantially higher medium intensity shares than rational choice across most demand scenarios and population compositions. In contrast to other behavioural types, Maximisers contribute comparatively little to medium intensity land management and are more strongly associated with high intensity land use and conservation-oriented management depending on demand conditions (Figure \ref{fig:es_per_type}). Interestingly, Environmental Implementers do not consistently exhibit the highest conservation shares across demand scenarios and population compositions. In most settings, Maximisers display higher conservation shares within type (Figure \ref{fig:selected_type_dynamics_lif}).

Behavioural composition not only affects aggregate outcomes, but also the ecosystem service provision of individual behavioural types (Figure \ref{fig:es_per_type}). Comparing per-agent ecosystem service provision between the North and Southwest compositions reveals substantial composition-dependent differences for some types. Averaged across demand scenarios, Maximisers exhibit the strongest differences in material ecosystem service provision between the two population contexts, whereas Traditionalists show the largest differences in non-material ecosystem service provision. In contrast, Social Satisfiers exhibit comparatively small composition-dependent differences in per-agent ecosystem service provision.

Aggregate system dynamics are substantially less variable than the dynamics of individual behavioural types. This suggests that heterogeneous behavioural compositions generate compensatory dynamics, where asynchronous behavioural responses of different types dampen aggregate fluctuations and thereby reduce temporal variability at the system level. More generally, behavioural heterogeneity reduces synchronised switching between land management practices, resulting in smoother aggregate dynamics and reduced oscillatory behaviour compared to rational choice.

\section{Discussion}

This study examined how empirically informed behavioural heterogeneity among forestry practitioners affects emergent land management patterns and ecosystem service provision in an agent-based land use model using a stylsied landscape. By separating decision-making types from realised land management practices, the model allows behavioural characteristics such as environmental attitudes, social norms, and behavioural inertia to influence land use decisions without assuming a direct one-to-one relationship between actor type and land use. The results show that behavioural heterogeneity changes both the temporal dynamics and spatial organisation of land management compared to a homogeneous rational choice benchmark. In particular, heterogeneous behavioural compositions dampen system-level fluctuations, increase the prevalence of medium intensity land management, and generate distinct ecosystem service outcomes depending on the behavioural composition. At the same time, the behaviour of individual decision-making types depends strongly on the population context in which they are embedded, demonstrating that behavioural type characteristics cannot be interpreted as direct predictors of land use outcomes.

\subsection{How behavioural heterogeneity shapes land use dynamics}

\subsubsection{Behavioural heterogeneity dampens synchronised responses}
A central finding of this study is that behavioural heterogeneity reduces immediate and synchronised responses to changing ecosystem service demand. Rapid adjustment of this kind is common in models based on rational-choice or optimisation assumptions and can produce large, coordinated shifts in land management at the system level. However, empirical land use change is often more incremental and heterogeneous \citep{lambin_land_2010, van_vliet_manifestations_2015}. This discrepancy has been identified as a weakness of integrated assessment models, which produce unrealistic outcomes when projecting future land use change \citep{turner_unprecedented_2018, perkins_toward_2023}. By assuming that land managers act as unconstrained economic optimisers with perfect information, such models overlook real-world adoption time-lags and socio-cultural constraints \citep{perkins_toward_2023}. Consequently, they project unprecedented and structurally implausible rates of land use change such as immediate, massive expansions of energy crops that vastly exceed any historical precedent of commodity crop expansion \citep{turner_unprecedented_2018}.  

In the CRAFTY setting, where multiple agents act independently and without perfect knowledge of other agents’ simultaneous decisions, synchronised immediate responses additionally lead to persistent alternation between relatively high conservation and relatively high intensification. These dynamics are especially pronounced in transitional areas where neither productive nor natural capital clearly dominates. In such areas, small changes in relative ecosystem service undersupply repeatedly shift the competitive advantage between management practices. Because many agents respond in the same way to these signals, local switches accumulate into system-level oscillations in land-management intensity shares and can prevent exact demand matching.  

This reflects a more general property of complex systems in which many agents respond identically to the same signal. Similar dynamics have been described in other systems. For example, models of honeybee thermoregulation show that colonies can fail to regulate temperature efficiently when all individuals respond at the same critical threshold, producing cycles of overcooling and overheating \citep{miller_complex_2009}. Conversely, top-down optimisation-based models do not show such systemic oscillations because they solve directly for exact macro-level equilibria, preventing targets from being exceeded \citep{brown_how_2021}. However, this structural feature relies on assuming total system coordination, which typically forces the exact target to be immediately reached \citep{turner_unprecedented_2018, brown_how_2021}. 

For heterogeneous behavioural compositions, differences in giving-in thresholds, environmental attitudes, perceived social norms, and behavioural inertia prevent all agents from changing management simultaneously. Instead, responses are distributed over time, which reduces synchronised switching, dampens aggregate oscillations, and leads to more gradual land use change.

\subsubsection{Behavioural diversity promotes intermediate land-management strategies}

Behavioural heterogeneity shifts land-use outcomes away from spatially polarised land
allocation and towards intermediate land-management strategies. While the rational choice
benchmark produces a stronger spatial separation between high-intensity land
management in areas with high productive capital and conservation-oriented management
in areas with high natural capital, behaviourally heterogeneous agents maintain higher
shares of medium-intensity land management, particularly in transitional zones where
productive and natural capital overlap. This model dynamic mirrors the underlying empirical reality of the surveyed European forestry practitioners. Across all five identified behavioural types, the highest proportion of respondents used medium intensity strategies (weak and strong Continuous Cover Forestry), outnumbering those practicing high-intensity clearcutting or inactive conservation management \citep{diana_feliciano_decision_2025}.  

This pattern can be related to the distinction between land sparing and land sharing \citep{green_farming_2005, phalan_reconciling_2011}, which was originally developed in the context of agriculture but has also been applied to forest systems \citep{edwards_landsharing_2014}. In a land-sparing strategy, production and
conservation are spatially separated, whereas land sharing combines production with
biodiversity conservation and non-material ecosystem service provision on the same land. In the model, the rational-choice benchmark most closely resembles a land-sparing logic,
while the higher prevalence of medium-intensity management under behavioural heterogeneity
resembles a land-sharing or multifunctional strategy, because it provides both
material and non-material services on the same land.

An important result is that the rational-choice benchmark does not always provide the closest match between ecosystem service supply and demand, even though rational agents respond directly to marginal demand signals. In some demand scenarios, especially when material demand is high but non-material demand remains present, the higher share of medium-intensity management under behavioural heterogeneity improves demand matching because it contributes to both service categories. This is particularly relevant in transitional zones, where intermediate management can make use of productive potential while maintaining some capacity for non-material service provision. In other scenarios, however, a stronger spatial separation between conservation areas and high-intensity management in productive areas produces higher total ecosystem service supply. This context dependence is consistent with the broader land-use literature, which likewise shows that the relative performance of sparing- and sharing-like patterns depends on landscape context, spatial scale, and socio-economic conditions \citep{fischer_land_2014,kremen_reframing_2015}. 

For socio-ecological modelling, this implies that intermediate land-management strategies
should not be treated merely as inefficient deviations from optimal land allocation, as commonly assumed in many land use models based on economic behaviour only.
Real land managers rarely optimise along a single objective. Their decisions often reflect
compromises between income, environmental values, local norms, perceived feasibility and existing practices \citep{tiebel_small-scale_2021,hugosson_objectives_2004}. Representing behavioural heterogeneity therefore allows the
model to capture plausible forms of multifunctional land management that are difficult
to reproduce under homogeneous rational-choice assumptions.

 \subsubsection{Regional behavioural compositions affect land use outcomes through different mechanisms}

Our experiments show that similar aggregate land use outcomes can emerge through different behavioural mechanisms. In the two tested heterogeneous behavioural compositions, the share of medium-intensity land management increases relative to the rational choice scenario. However, the mechanisms producing this outcome differ. In the behavioural composition corresponding to forestry practitioners' profiles in Southwest Europe, the higher prevalence of agents influenced by social norms supports the adoption of medium-intensity management as a socially acceptable compromise between more intensive and more conservation-oriented practices. In the behavioural composition representing the decision-making types of North European forestry practitioners, by contrast, medium-intensity management emerges through a combination of mechanisms, particularly the persistence of existing practices among behavioural types characterised by stronger behavioural inertia.

This suggests that aggregate land use patterns alone may hide substantial differences in the processes that generate them. The same broad outcome may reflect either adaptive social coordination or path-dependent persistence. This distinction has broader implications for the interpretation of land use and ecosystem service outcomes, because similar observed landscapes may differ in their capacity for future change depending on the behavioural mechanisms that maintain them. 

In a landscape in which intermediate management is sustained by social responsiveness, the adaptivity of the system depends highly on the homogeneity of the neighbourhood. In heterogeneous neighbourhoods socially responsive agents are exposed to multiple practices and can quickly align with emerging local patterns. On the other side, in landscapes, where a dominant practice is already strongly clustered, the same norm-following mechanism can reinforce the prevailing behaviour and make change more difficult.
By contrast, a landscape in which intermediate management is sustained by inertia may be less adaptable independently on the local social context.

For ecosystem service governance, these findings suggest that the effectiveness of interventions may depend not only on the current land use pattern, but also on the behavioural composition of land managers. Behavioural and policy studies show that policy instruments are more effective if their design considers the characteristics and motivations of the target population \citep{pineiro_scoping_2020, lee_assessment_2019, schimmelpfennig_cultural_2025}.
Policies that rely on demonstration effects, peer learning, or local norm formation may be more effective where socially responsive agents dominate \citep{sparkman_how_2021,gronow_policy_2021,ingram_enabling_2018}. Contexts characterised by stronger inertia, in contrast, may require longer-term incentives, institutional support, or measures that reduce the perceived costs of transition \citep{pineiro_scoping_2020, dessart_behavioural_2019}. 
This is relevant for socio-ecological modelling because it shows that behavioural composition can influence both present ecosystem service provision and the future adaptability of land systems.

\subsubsection{Context dependence complicates the prediction of behavioural-type outcomes}

The results also show that behavioural types do not have fixed effects on land management outcomes. Their realised behaviour depends on the demand scenario and on the population composition in which they are embedded. This context dependence complicates the prediction of behavioural-type outcomes from their decision-making profiles alone.

Social Satisfiers illustrate how sensitivity to social norms can generate strong short-term dynamics. Because they respond to neighbouring behaviour, they react quickly to the initially random distribution of management practices. This leads to rapid early increases in medium intensity land management and to the formation of local clusters, particularly in transitional zones. However, their longer-term behaviour depends on the practices adopted by surrounding agents. Social Satisfiers therefore do not simply represent a fixed tendency towards medium intensity management; rather, they amplify and stabilise locally emerging patterns.

Environmental Implementers and Maximisers provide more counterintuitive examples. Environmental Implementers have strong pro-environmental attitudes and initially adopt conservation-oriented management rapidly. However, widespread extensification can reduce material ecosystem service supply, creating demand pressure for subsequent intensification. As a result, Environmental Implementers do not consistently produce the highest long-term conservation shares. Conversely, Maximisers have a productivist orientation but can adopt relatively high shares of conservation-oriented management in some settings. This occurs because they remain responsive to demand signals and allocate land more closely according to capital suitability. When non-material ecosystem services are undersupplied, conservation can become competitive even for agents with negative environmental attitudes.

These findings highlight the value of distinguishing decision-making types from land management practices. A behavioural type such as an Environmental Implementer should not be interpreted as a direct proxy for conservation land use, just as a Maximiser should not be assumed to always produce intensive management. Behavioural types define tendencies, but realised outcomes emerge from the interaction between behavioural parameters, social context, landscape suitability, and ecosystem service demand feedbacks. This is particularly important for using empirical typologies in agent-based models. Decision-making typologies provide valuable information on differences in beliefs, objectives, and normative orientations, but their land use implications depend on the socio-ecological system in which agents act \citep{ficko_european_2019, diana_feliciano_decision_2025}. Land managers often possess highly diverse objectives and values \citep{tiebel_small-scale_2021, westin_forest_2023} even when current conditions constrain them to manage in similar ways, meaning a significant desire for alternative management may remain latent until the broader context shifts.

 \subsection{ Realism and limitations of the behavioural parametrisation}
 
The translation from empirical typology to model parameters necessarily involves interpretation. The empirical typology provides information on behavioural, normative, and control beliefs, but these characteristics cannot be mapped directly onto model parameters such as environmental attitude, social norm weight, behavioural inertia, or giving-in thresholds. The resulting parametrisation should therefore be understood as a theoretically informed operationalisation rather than as a direct empirical measurement of decision rules.  

This interpretive step may also help explain some mismatches between the typology narrative and simulated behaviour. For example, Environmental Implementers are described in the empirical typology as actors who seek to operationalise environmental objectives and who engage with market mechanisms when these support such objectives. In the model, however, they do not always extensify more strongly than Maximisers under conditions of high non-material ecosystem service demand. One possible reason is that the current parametrisation does not explicitly represent forestry networks, institutional actors, or policy instruments such as subsidies and grants, even though these may be particularly important for this group. This suggests that future refinements of the behavioural layer could improve the representation of how institutional and policy contexts shape the behaviour of different practitioner types.

\subsection{Scope of the study and future research}

The model is deliberately stylised in several respects. It uses a simplified landscape, a reduced set of land-management intensities, and highly aggregated ecosystem service categories. In addition, the regional behavioural compositions are used to represent empirically observed differences in decision-making structure, but not to reproduce the actual geography or institutional context of Northern or Southwestern Europe. These design choices make it possible to isolate the effects of behavioural heterogeneity and to interpret the resulting dynamics more clearly. The results should therefore be understood as theoretical insights into how socio-psychological diversity shapes land-use dynamics and ecosystem service provision, rather than as applied predictions for specific regions. At the same time, the framework provides a basis for future development towards more context-specific simulations.

A further simplifying assumption concerns the spatial assignment of behavioural types. Although agents are distributed according to empirically observed regional shares, behavioural types are assigned randomly across the landscape. This may not reflect real-world
social and spatial patterns. For instance, land managers with similar behavioural profiles, preferences, and values may be spatially clustered. Because social norms in the model operate through local neighbourhoods, the spatial distribution of behavioural types and initial practices can influence the strength and timing of norm-driven dynamics. More clustered configurations, for example, could reinforce prevailing practices and generate stronger local lock-in effects, particularly for strongly norm-sensitive types. Future work can apply more realistic initial spatial distributions of behavioural types.

\section{Conclusion}

This study demonstrates that empirically informed behavioural heterogeneity among forestry practitioners can substantially shape emergent land use dynamics and ecosystem service outcomes as well as individual behavioural dynamics in agent-based land use models. By translating a survey-based behavioural typology into cognitive parameters within CRAFTY, we show that socio-psychological differences among land managers matter not only for the direction of land use change, but also for its timing and stability. Compared with a rational-choice benchmark, heterogeneous behavioural compositions dampen synchronised switching, reduce oscillatory dynamics, and increase the prevalence of medium-intensity management.

The results further show that aggregate outcomes depend on both regional behavioural composition and the context-dependent behaviour of individual types. Behavioural composition influences ecosystem service provision through multiple mechanisms, while the same behavioural type can contribute to different land use outcomes depending on demand conditions and the surrounding population. This has important implications for socio-ecological modelling and governance: empirical behavioural typologies should not be treated as direct proxies for performed land use practices, and interventions may work differently depending on the behavioural composition of the target population.

Overall, the study provides a clear proof of concept that socio-psychological diversity should be represented explicitly in agent-based land use models, especially when these are used to explore sustainability transitions and ecosystem service provision. Future work should test these mechanisms in real landscapes and under dynamic socio-ecological conditions.

\section*{Code availability}

The model code and documentation (ODD protocol) have been archived in the CoMSES Computational Model Library for peer review. The model will be made publicly available through CoMSES upon acceptance of the manuscript.

\section*{Acknowledgements}

This research was supported by the ForestPaths project (Co-designing Holistic Forest-based Policy Pathways for Climate Change Mitigation), funded under the European Commission Horizon Europe Research and Innovation Action (Grant Agreement No. 101056755).

We are especially grateful to Florencia Franzini for her invaluable support in working with the empirical typology developed in the ForestPaths deliverable “Decision rules, parameters, and narratives for modelling”. She provided essential guidance in interpreting the typology and its indicators, and her insights were crucial for the parametrisation of behavioural types through detailed discussions.

We also thank Joanna Raymond for her helpful contributions to discussions on the typology and its parametrisation, as well as for her input on the allocation of behavioural type compositions across different regions.

\newpage
\appendix
\renewcommand{\thefigure}{A\arabic{figure}}
\setcounter{figure}{0}
\renewcommand{\thetable}{A\arabic{table}}
\setcounter{table}{0}
\section{Supplementary Figures}
\label{appendix_A}

\begin{figure}[H]
\centering
\subfigure[Productive capital distribution]{
  \includegraphics[width=0.45\linewidth]{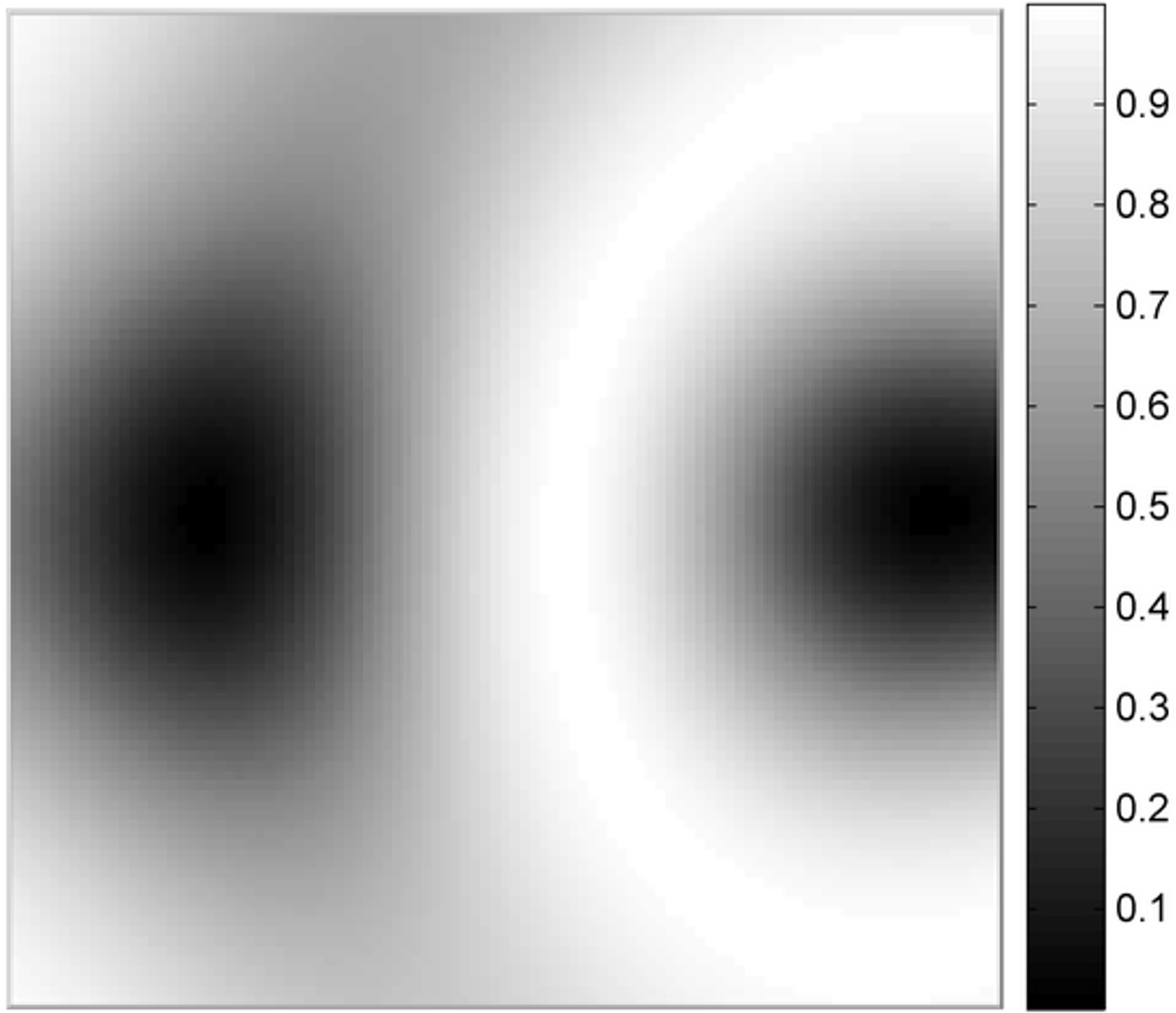}
  \label{fig:crop}
}
\hfill
\subfigure[Natural capital distribution]{
  \includegraphics[width=0.45\linewidth]{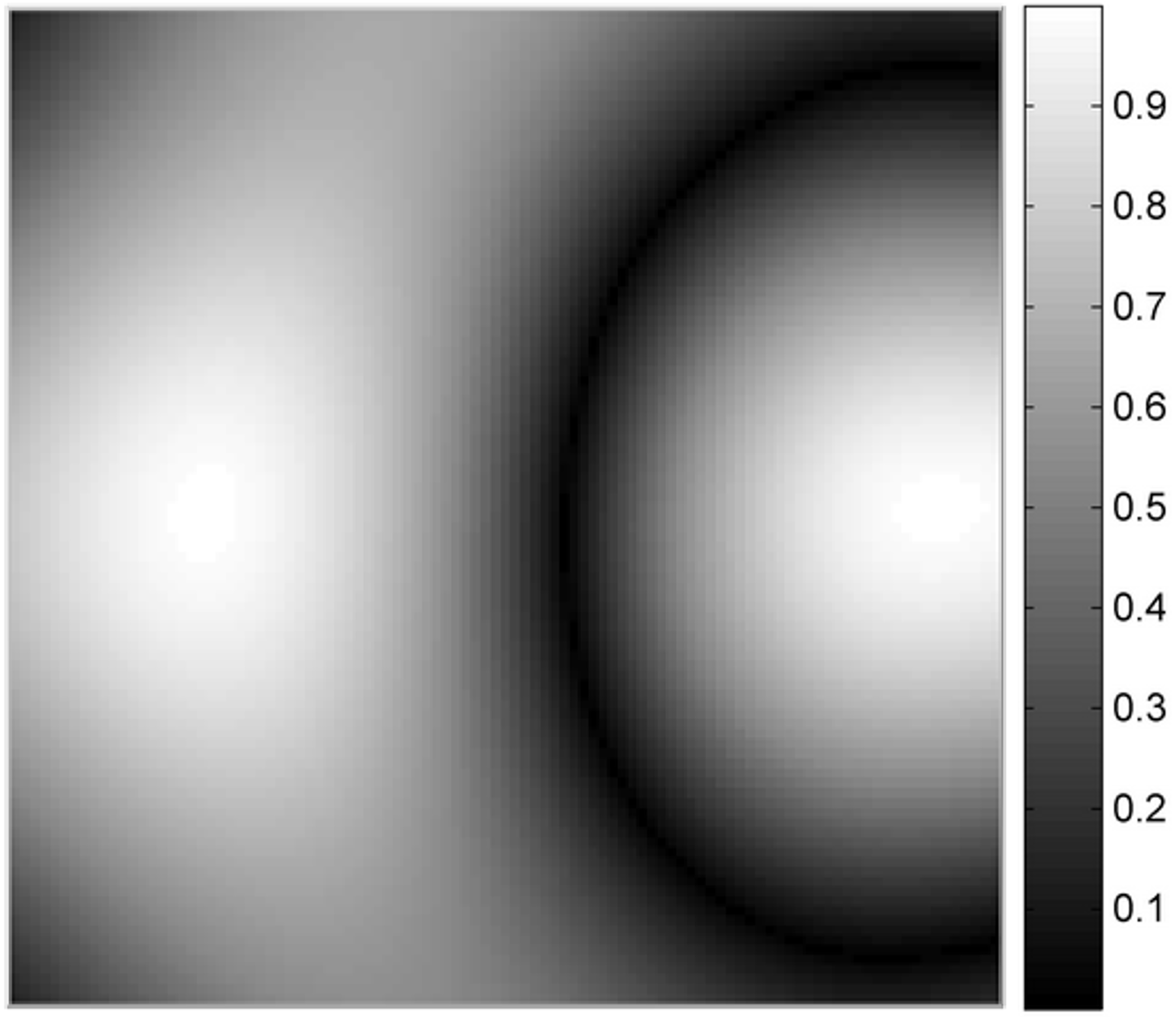}
  \label{fig:natural}
}
\caption{Productive and natural capital distribution across the modelled landscape, representing the land's suitability to produce material and non-material ES, respectively. Lighter colours indicate higher capital levels.}
\label{fig:capitals}
\end{figure}

\section{Supplementary Tables}
\label{appendix_B}

\begin{table}[h]
\centering
\caption{Global model parameters held constant across all simulation experiments.}
\label{tab:global_parameters}
\begin{tabularx}{\linewidth}{l c X}
\toprule
Parameter & Value & Description \\
\midrule
Critical mass threshold & 0.5 & Share of neighbours required to trigger social norm influence \\
Teleconnections & 0 & No long-distance social ties included \\
Neighbourhood radius & 2 & Spatial radius for social norm calculation \\
Logistic steepness & 10 & Controls sharpness of behavioural response threshold \\
Initial intensity shares & 1/3 each & Equal shares of high, medium, and conservation management \\
\bottomrule
\end{tabularx}
\end{table}

\begin{table}[H]
\centering
\small
\renewcommand{\arraystretch}{1.15}
\setlength{\tabcolsep}{4pt}

\caption{Empirical indicators underlying the behavioural typology. Entries summarise the direction and strength of each aggregated empirical indicator for each behavioural type. The symbols denote strong negative (-- --), negative (--), ambivalent (+/--), positive (+), and strong positive (++). The six indicators are aggregated from multiple empirical belief items: regulating ecosystem service objectives (RES), income objectives (Income), amenity objectives (Amenity), forestry networks (Network), societal referents (Society), and market mechanisms (Market). Society refers to societal referents such as neighbours, friends, and the general public; Network refers to forestry networks, including local and national forest authorities and forest owner associations; and Market refers to market mechanisms such as subsidies and grants, market schemes, and credit or loans. Full details on the underlying survey items and their empirical derivation are provided in \citet{diana_feliciano_decision_2025}.}
\label{tab:typology_matrix}

\begin{tabular}{l c c c c c c}
\toprule
 & RES & Income & Amenity & Network & Society & Market \\
\midrule

\textbf{Env. Conscious Passives} 
& + & +/-- & -- & +/-- & +/-- & -- -- \\

\textbf{Env. Implementers} 
& + & -- -- & +/-- & ++ & -- -- & + \\

\textbf{Traditionalist} 
& + & +/-- & +/-- & -- -- & + & +/-- \\

\textbf{Maximisers} 
& -- -- & + & -- -- & +/-- & -- & ++ \\

\textbf{Social Satisfiers} 
& +/-- & +/-- & + & + & + & +/-- \\

\bottomrule
\end{tabular}

\end{table}

\section{Interpretative translation of empirical indicators into model parameters}
\label{appendix_C}
The behavioural parameterisation was based on an interpretative translation of the
empirical typology rather than on direct statistical estimation. The empirical indicators
reported in Table~A2 describe the direction and relative strength of different behavioural
characteristics for each practitioner type using ordinal categories ranging from strongly
negative to strongly positive. These indicators were used as qualitative coding evidence
for assigning discrete model parameter values, not as quantitative measurements of
parameter magnitudes.

The environmental attitude parameter \(A_{\alpha}\) was derived from the relative strength
of regulating ecosystem service objectives compared with income objectives. This reflects
the interpretation of \(A_{\alpha}\) as a signed orientation towards conservation-oriented
versus production-oriented management. Types with stronger regulating ecosystem service
objectives than income objectives were assigned more positive attitude values, whereas
types with stronger income objectives than regulating ecosystem service objectives were
assigned lower or negative values.

The social norm weight \(w_{\alpha}\) was derived from the society indicator relative to the
attitude component. The society indicator captures the importance assigned to societal
referents such as neighbours, friends, and the general public. Since \(w_{\alpha}\) represents
the relative weight of social norms compared with attitude in the decision layer, types
with stronger society indicators relative to their attitude component were assigned higher
social norm weights.

The upper limit of the giving-in threshold \(L_{\alpha}\) was informed by income objectives
and market mechanisms relative to other motivations. This parameter was interpreted as
capturing how readily agents respond to competitive or demand-related incentives rather
than remaining guided by other motivations. Types with stronger income objectives and
market orientation were therefore interpreted as more responsive to competitiveness
signals, while types whose empirical profile emphasised non-economic objectives or weak
market orientation were assigned higher thresholds.

Behavioural inertia \(\lambda_{\alpha}\) could not be derived from a single empirical indicator in
Table~\ref{tab:typology_matrix}. It was therefore assigned from the qualitative descriptions of the practitioner
types. Types described as passive, less active, or reluctant to change management were
assigned higher inertia values, whereas types described as more active or
implementation-oriented were assigned lower inertia values.

Amenity objectives and forestry network indicators were not directly translated into
separate parameters. This is because the current behavioural model does not include
distinct mechanisms for amenity-oriented objectives and does not represent forest owner associations and local and national forestry authorities. These indicators were therefore used only as contextual information
when checking whether the resulting parameterisation remained consistent with the
overall type descriptions.
\\
\\
\textbf{Environmentally Conscious Passives}
\\
Environmentally Conscious Passives show little interest in deriving personal utility from forest management and attach limited importance to social norms or market mechanisms when making management decisions. As a result, few drivers motivate behavioural change, which is why this group is characterised as “passive”. In the model, this disposition is represented by a high maximum giving-in threshold combined with behavioural inertia. Unlike the passive owner archetype described in parts of the forestry literature, however, this group also places strong emphasis on environmental objectives, including improving soil and water quality and mitigating climate change and forest disturbances. We reflect this orientation by assigning a higher weight to environmental attitudes than to social norms and by specifying a positive environmental attitude parameter for this type.
\\
\\
\textbf{Environmental Implementers}
\\
Environmental Implementers place strong importance on environmental objectives while assigning little importance to income considerations and to social norms from neighbouring practitioners. In the model, this orientation is represented by assigning no weight to social norms and a strongly positive environmental attitude.

At the same time, this group places considerable importance on forestry networks, such as local and national forest authorities and forest owner associations, and recognises the role of market mechanisms such as subsidies and grants. This suggests that Environmental Implementers value the knowledge and guidance provided by professional networks and the financial resources required to carry out forest management activities. They appear motivated to operationalise their environmental objectives while acknowledging the institutional and economic conditions necessary to implement them.

Because income objectives are of little importance but this group is willing to actively implement management changes, we assign a slightly lower maximum giving-in threshold than for Environmentally Conscious Passives and assume no behavioural inertia. In our model, forestry networks and market mechanisms are not represented explicitly. Instead, we reflect the willingness of this group to act when opportunities align with their environmental objectives through a high environmental attitude combined with a moderately high giving-in threshold, allowing agents to respond to demand signals that are consistent with their environmental values.
\\
\\ \textbf{Traditionalists}
\\
Traditionalists are characterised by a strong emphasis on environmental objectives, a high reliance on social norms from neighbouring practitioners, and relatively low importance placed on income objectives. These characteristics correspond closely to the Traditionalist archetype commonly identified in the forestry literature.

In the model, we translate these traits into a positive environmental attitude and a balanced weighting between environmental attitudes and social norms. In addition, we assign a high maximum giving-in threshold, meaning that both social norms and environmental attitudes strongly shape decision-making while competitive responses to marginal demand signals play a smaller role.

While placing considerable value on societal norms related to local traditions, Traditionalists see broader referent groups such as forestry authorities as rather unimportant. To capture this tendency to value local traditions, together with the only moderate importance attributed to market mechanisms, we introduce a moderate level of behavioural inertia for this type.
\\
\\
\textbf{Maximisers}
\\
Maximisers place very strong emphasis on income objectives, assign importance to market mechanisms, and see environmental and amenity objectives as highly unimportant, reflecting a strongly productivist orientation. In the model, we translate this profile into a low maximum giving-in threshold, no behavioural inertia, and a negative environmental attitude, corresponding to a preference for more intensive forest management. This parameter combination makes the type highly responsive to marginal demand signals while biasing decisions towards the provision of material ecosystem services, such as timber, rather than regulating or cultural services associated with more conservation-oriented land management. In addition, Maximisers place no importance on social norms, which we represent by assigning no weight to descriptive norms from neighbouring forestry practitioners in our model.
\\
\\
\textbf{Social Satisfiers}
\\
Social Satisfiers are characterised primarily by their sensitivity to social norms, especially societal opinion, while also assigning some importance to the views of forestry networks. At the same time, they remain ambivalent towards income objectives, environmental objectives, and market mechanisms. We translate this configuration into the model by assigning a high weight to descriptive norms from neighbouring forestry practitioners, a neutral environmental attitude, a moderate maximum giving-in threshold, and no behavioural inertia.
\newpage

\bibliographystyle{jasss}

\bibliography{references} 
\end{document}